\documentclass[pra,superscriptaddress]{revtex4}

\usepackage{graphicx,color}
\usepackage{amsmath}
\usepackage{amssymb}
\usepackage[normalem]{ulem}
\usepackage[colorlinks=true,citecolor=blue]{hyperref}
\def\ner{\boldsymbol}

\usepackage{tikz}
\usetikzlibrary{patterns}

\newcommand{\ket}[1]{|{#1}\rangle}
\newcommand{\bra}[1]{\langle{#1}|}

\newcommand{\Up}{{\uparrow}}
\newcommand{\Dn}{{\downarrow}}
 \newcommand\beq            {\begin{equation}}
  \newcommand\eeq          {\end{equation}}
 \newcommand\bwt         {\begin{widetext}}
 \newcommand\ewt         {\end{widetext}}

\usepackage{times}
\usepackage{amssymb}
\usepackage{amsfonts}
\usepackage{amsmath}
\usepackage{graphicx}
\usepackage{bm}
\usepackage{enumerate}

\def\tfract#1/#2{{\textstyle{\raise0.8pt\hbox{$\scriptstyle#1$}\over%
\hbox{\lower0.8pt\hbox{$\scriptstyle#2$}}}}}
\def\mezzo{\tfract 1/2 }

\def\radi2k{\tfract 1/{\sqrt {2k}} }
\def\der{\partial}
\def\downnormalfill{$\,\,\vrule depth4pt width0.4pt
\leaders\vrule depth 0pt height0.4pt\hfill\vrule depth4pt width0.4pt\,\,$}
\def\WT#1{\mathop{\vbox{\ialign{##\crcr\noalign{\kern3pt}
      \downnormalfill\crcr\noalign{\kern0.8pt\nointerlineskip}
      $\hfil\displaystyle{#1}\hfil$\crcr}}}\limits}

\def\be{\begin{equation}}
\def\ee{\end{equation}}
\def\bes{\begin{equation*}}
\def\ees{\end{equation*}}
\def\bea{\begin{eqnarray}}
\def\eea{\end{eqnarray}}
\def\beas{\begin{eqnarray*}}
\def\eeas{\end{eqnarray*}}
\def\ba{\begin{array}{rcl}}
\def\ea{\end{array}}
\def\der{\partial}
\numberwithin{equation}{section}
\def\go{\leavevmode \raise.3ex\hbox{$\scriptscriptstyle \langle\!\langle\!  $}%
~\ignorespaces}

\def\gf{\relax \ifhmode \unskip~\else \leavevmode \fi \raise.3ex\hbox{$\! \scriptscriptstyle\rangle\!\rangle\, $}}
\begin{document}

\author{Luca Lepori}
\affiliation{Dipartimento di Scienze Fisiche e Chimiche, Universit\`a dell'Aquila, via Vetoio, I-67010 Coppito-L'Aquila, Italy.}
\affiliation{INFN, Laboratori Nazionali del Gran Sasso, Via G. Acitelli, 22, I-67100 Assergi (AQ), Italy.}

\author{Michele Burrello}
\affiliation{Niels Bohr International Academy and Center for Quantum Devices, University of Copenhagen, Juliane Maries Vej 30,
2100 Copenhagen, Denmark.}

\author{Enore Guadagnini}
\affiliation{Dipartimento di Fisica E. Fermi, Universit\`a  di Pisa, Largo B. Pontecorvo 3, 56127 Pisa, Italy.}
\affiliation{INFN, Sezione di Pisa, Largo B. Pontecorvo 3, 56127 Pisa, Italy.}

\title{ Axial anomaly in multi-Weyl and triple-point semimetals}

\begin{abstract}

We derive  the  expression of the abelian axial anomaly  in  the so-called multi-Weyl and triple-point crossing semimetals.  No simplifying restrictions are assumed  on the symmetry of the spectrum.  Three different  computation methods are considered:  the perturbative quantum field theory procedure which is based on the evaluation of the one-loop Feynman diagrams,  the Nielsen-Ninomiya  method, and  the Atiyah-Singer index argument.  It is shown that the functional form of the axial anomaly does not depend on the Lorentz  symmetry,  but it is determined by the gauge structure  group.  We discuss the stability of the anomaly -- stemming from the quantization of the anomaly coefficient -- under smooth modifications of the lagrangian parameters. 
  
\end{abstract}

\maketitle

\section{Introduction}

Anomalies are known since long ago in the context of quantum field theory \cite{IZ,peskin,weinberg}.  In the construction of a quantum field theory based on a classical lagrangian,  it may happen that a certain symmetry of the classical action cannot be preserved at the quantum level; when this happens, the symmetry is called anomalous. 
 A well known  example of anomaly is found when chiral fermions are minimally coupled with gauge fields; in this case, some of the  symmetries acting on the fermion fields may be broken at the quantum level  because of the so-called chiral anomalies  \cite{adler1969,bell1969,S,WZ,BZ}.  In elementary particle physics, the experimental consequences of the flavour  chiral anomalies have  been described for instance in  \cite{adler1969,bell1969,WZ,THK}.

In the last decade it has been realized that the study of anomalies in field theories is also a fundamental tool for the effective description of topological phases of matter such as topological insulators and superconductors \cite{hasankane,zhang11}. In this scenario it has been shown that a field theoretical approach accounting for chiral and gravitational anomalies allows us to characterize the peculiar transport properties of topological materials resulting from the coupling to electromagnetic fields or temperature gradients \cite{ryu2012,burkov2012}. In particular, anomalies provide a natural description for phenomena like  
the surface Hall conductance, which are related to the gapless surface modes of these gapped systems and constitute a useful tool for their classification.

Since the work by Nielsen and Ninomiya \cite{nielsen1983} it has been known that also gapless models can enjoy similar topological responses under external electromagnetic fields, thus actualizing the effects of quantum anomalies. Only recently, however, similar topological gapless phases of matter have gained a considerable attention and have been experimentally realized in solid state materials \cite{lv15a,lv15b,su2015,xu2015a}. 
The main example is provided by Weyl semimetals \cite{wan11,balents11}, which constitute a remarkable embodiment of the Dirac theory for massless fermions.  In the presence of magnetic fields, they display transport properties which are dictated by the corresponding chiral anomaly \cite{burkov2012} and have been studied in recent works  \cite{vis2014,xiong2015,huang2015,zhang2016,liexp2016}. 
These three dimensional systems host pairs of inequivalent  and isolated Weyl points (or Weyl nodes) in the Brillouin zone. These are points where two energy bands touch each-other, with a linear dispersion that determines the appearance of a cone. The Weyl nodes appear always in pairs for lattice models \cite{nielsen1981} and can be separated in momentum space by breaking the space inversion or time reversal canonical symmetries \cite{wan11,balents11,lepori2015}.
In their neighborhood, thus for energies close to the band touching points, the fermionic quasiparticles display a linear dispersion law and their dynamics can be effectively described by a Weyl hamiltonian involving a pair of cones with opposite ``chirality''. 
The appearance of Weyl points in pairs has a topological origin \cite{nielsen1981,rothe}; moreover it also implies (if space-rotational symmetry is present close to the nodes) the emergence of an effective Lorentz covariance, characterizing the low-energy dynamics of the Weyl semimetal, the chemical potential is assumed to coincide with the energy of the Weyl nodes.
 
Weyl semimetals present indeed non-trivial topological features, exactly due to the Weyl nodes which constitute a monopole source for the Berry connection of the bands \cite{volovik2003}. This topology manifests in the presence of gapless chiral modes, exponentially localized on the surfaces of these systems. At the energy of the Weyl nodes, such surface states identify lines in the momentum space connecting the projections on the surface Brillouin zone of two inequivalent nodes; these surface modes are therefore dubbed Fermi arcs \cite{wan11}.
 Weyl semimetals and Fermi arcs have been realized and detected in various compounds, including for instance tantalum arsenide \cite{lv15a,lv15b,su2015},  niobium arsenide  \cite{xu2015a}, bismuth trisodium \cite{xu2015b},  and tantalum phosphide \cite{xu2017}. Their implementation, however, was not limited to solid-state materials only, but it also encompasses other platforms, for example in photonic \cite{lu15,chen16,noh17} and phononic \cite{li18} crystals.

Quite recently, Weyl semimetals  (which we will dub single-Weyl semimetals in the following, to avoid confusion) have been shown to admit notable generalizations in the so-called multi-Weyl semimetals \cite{bernevig2012}, lattice systems displaying
isolated band touching points, similar to the Weyl nodes, but where the dispersion law is linear along one space direction only and
grows with a higher power of the momenta  along the other two directions. Double-Weyl nodes have been predicted in a number of rare-earth compounds
\cite{fang2011,shivamoggi2013,yao2015,yao2016,fiete2016,hasan2016}, in ultracold-atoms set-ups \cite{lepori2016,mai2017,lang2017} and in photonic crystals \cite{chen2016}.

The dispersion relation of the multi-Weyl points implies a breaking of the effective (low-energy) Lorentz covariance, which characterizes instead the single-Weyl cones. Despite the absence of the Lorentz symmetry,  it has been argued \cite{roy2015,li2016,shen2017} that, for particular values of the lagrangian parameters, the axial anomaly assumes  the standard functional form that one derives in Lorentz invariant theories. 
The anomalous Hall conductivity which characterizes multi-Weyl materials does not depend indeed on the effective Lorentz covariance of the system, which is always violated when one takes into account the band dispersion. The latter property has been exemplified by even more exotic topological semimetals, characterized by the simultaneous merging of multiple energy bands, such that they can be interpreted as models with a modified spin-statistics relationship \cite{bradlyn2016}. A notable example involves triple-point semimetals \cite{zhu2016,fulga2017,hu2017}, recently realized \cite{ding2017} in molybdenum phosphide, where three bands  cross in a triply-degenerate point with a multiple topological charge. A summary of some of the main features of these topological semimetals is provided in Sec. \ref{sec:base}.

The main  scope of the present work is the  computation  of  the abelian anomalies  related to  suitable global transformations of  anticommuting  fields,  which appear in the lagrangian models describing double-Weyl, triple-Weyl and triple-point lattice systems around the nodal points.  These anomalies are connected   with the so-called axial transformations acting on the fermion fields in the presence of  an abelian vector potential  $A_\mu (x)$  (see \cite{shen2017} and references therein).  An overview of our results is presented in section~\ref{overview}.  

No simplifying restrictions are assumed  on the symmetry of the spectrum of the various models.   The considered  low energy lagrangians are not Lorentz invariant; indeed,  the relevant differential operators acting on the fermionic variables are not necessarily  described by homogeneous polynomials of the covariant  derivatives,  and may contain  dimensioned parameters in front of them.  Let us recall that the standard results \cite{adler1969,bell1969,S,WZ,BZ} concerning the axial anomaly have been obtained in the presence of  Lorentz invariance. So the computation of the axial anomaly  in the case of models which are not Lorentz invariant presents original aspects. Therefore,  in order to make our article self-contained, in sections~\ref{general}-\ref{index} we describe in detail our anomaly computations in the case of the double-Weyl semimetal model; the   triple-Weyl and triple-point models are examined  subsequently.  

Three different   methods for the derivation of the anomaly are considered:    the perturbative quantum field theory procedure \cite{adler1969,bell1969,S,WZ,BZ} which is based on the evaluation of the one-loop Feynman diagrams,  the Nielsen-Ninomiya  method \cite{nielsen1983}, and  the Atiyah-Singer index argument \cite{AS,Nakahara}. The mutual consistency of these methods is illustrated. The general features of the  perturbative approach  are described in section~\ref{general}, where the relationship between the chiral anomaly and the axial anomaly is produced. 
The perturbative computations of the chiral anomaly for the double-Weyl model  are contained in section~\ref{perturb}, the Nielsen-Ninomiya procedure is presented in section~\ref{NN-procedure}, and the  anomaly derivation by means of the Atiyah-Singer index argument is contained in section~\ref{index}.   The axial anomaly for the triple-Weyl and the triple-point  models are derived in section-\ref{3-Weyl} and section~\ref{3-point} respectively. In these cases, the perturbative approach is rather arduous or affected by singularities; therefore only the Nielsen-Ninomiya and Atiyah-Singer methods are considered.  The quantization of the multiplicative anomaly coefficient is discussed in section~\ref{quantization}, which also contains a few comments on the structure of the obtained anomaly expressions, {\color{black} as well as on their stability  {\color{black} under smooth} perturbations of the lagrangian parameters}.   {\color{black} The effects  on the axial anomaly of modifications of the chemical potentials are examined in detail in  section~\ref{Stability}, where we also illustrate certain peculiar features of the anomaly computation in the case of the triple-point model.} Finally, our conclusions are presented in section~\ref{conclus}. 

{\color{black} \section{Basic features of topological semimetals} \label{sec:base}

The prediction and discovery of {\color{black} symmetry-protected topological matter  is considered one of the crucial achievements of the theory of condensed matter in the last decades. The most prominent example in this set is given by   topological insulators \cite{hasankane,zhang11}. These are gapped materials,  non-interacting or weakly interacting,} which present gapless modes localized on their surface or edges and responsible for their transport properties. These surface modes, in general, maintain their gapless and localized nature as long as certain discrete symmetries of the system are preserved or a phase transition through a critical point of the bulk occurs. Topological insulators can be efficiently described in terms of non-interacting lattice models and the existence of their gapless edge or surface modes can be deduced, in general, by topological indices, such as Chern or winding numbers, characterizing their energy bands in the Brillouin zone of the lattice.

The study of topological features in condensed matter materials extended rapidly to gapless systems (see, for example, the reviews \cite{chiu16,Armitage18}), namely topological metals and semimetals. A semimetal is a material that presents two partially filled energy bands at its chemical potential; this implies that there are at least two energy bands overlapping in energy. The limiting case is provided by energy bands with a discrete set of band touching points in the Brillouin zone and the chemical potential lying exactly at the energy of these points. The most typical  example of topological semimetal is the Weyl semimetal that belongs indeed to this case: a Weyl semimetal is a  three-dimensional material with two bands touching with a linear dispersion along all the directions in an even number of points in the Brillouin zone (see Fig. \ref{fig:disp}($a$)).{ \color{black} These gapless points are robust against any small translational-invariant perturbation of the system, as long as they lie at different momenta of the Brillouin zone. Any pair of these nodes with opposite chiralities can be efficiently described in terms of two decoupled Weyl hamiltonians; therefore,}
for energies close to the band-touching points (thus small temperatures and chemical potentials close to the Weyl points) these systems are characterized by an emerging Lorentz invariance; increasing or decreasing the chemical potential, instead, the non-trivial dispersion of the bands become relevant to determine the physical properties of the system. 

{\color{black} The linear-dispersing single-Weyl points correspond to unitary monopoles of the Berry curvature calculated on the two touching energy bands \cite{volovik2003}. These points} do not exhaust the possible topological nodes among different bands: in the presence of additional symmetries (e.g. rotational crystal symmetries) it is possible to engineer materials displaying multi-Weyl points \cite{bernevig2012}, which correspond to higher monopoles of the Berry curvature, at the price of giving away the emergent Lorentz invariance.
In particular, double-Weyl and triple-Weyl points are stabilized in physical material by the discrete rotational symmetries $C_4$ and $C_6$ and, in general, they do not display a full rotational SO(2) symmetry. Double-Weyl points are characterized by a quadratic dispersion along two directions ($k_1$ and $k_2$ in Fig.\ref{fig:disp}($b$)) and by a linear dispersion in the third direction. Triple-Weyl points, instead, display a cubic dispersion along two directions and a linear dispersion along the third one (in Fig. \ref{fig:disp} ($c$) an example with linear dispersion along $k_3$ and cubic dispersion along $k_1$ is displayed).
The triple-Weyl case is the one with the highest power in the dispersion law allowed by point group symmetries \cite{bernevig2012} in spatial dimensions equal or lower than three; by including multi-component fermions with additional symmetries, however, it is possible to engineer band-touching points with even larger dispersion powers and  monopole charges.
\begin{figure}[ht]
\includegraphics[width=.48\textwidth]{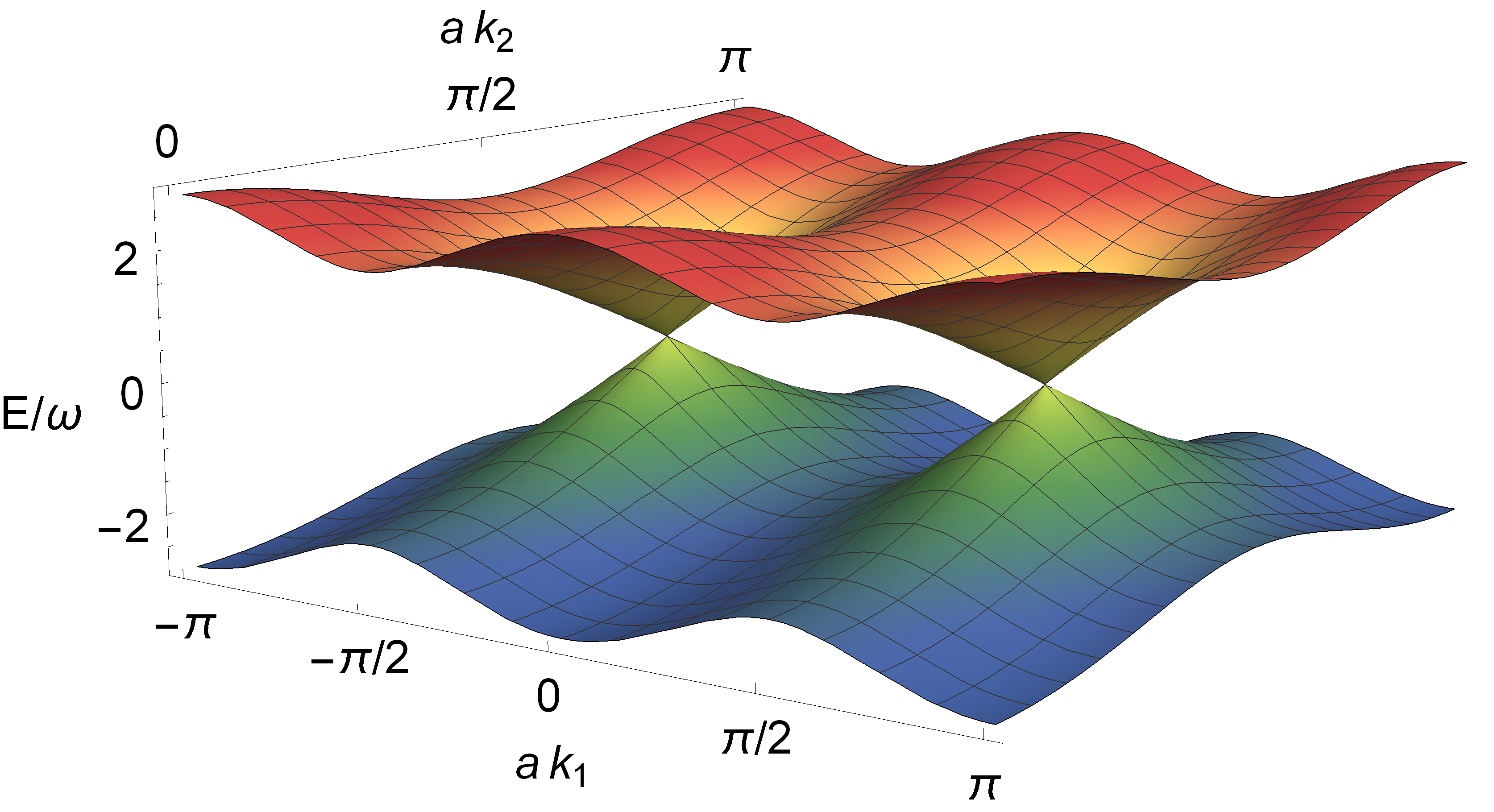}
	\llap{
  \parbox[b]{5cm}{$(a)$\\\rule{0ex}{3.9cm}
  }}\hfill
  	\includegraphics[width=.48\textwidth]{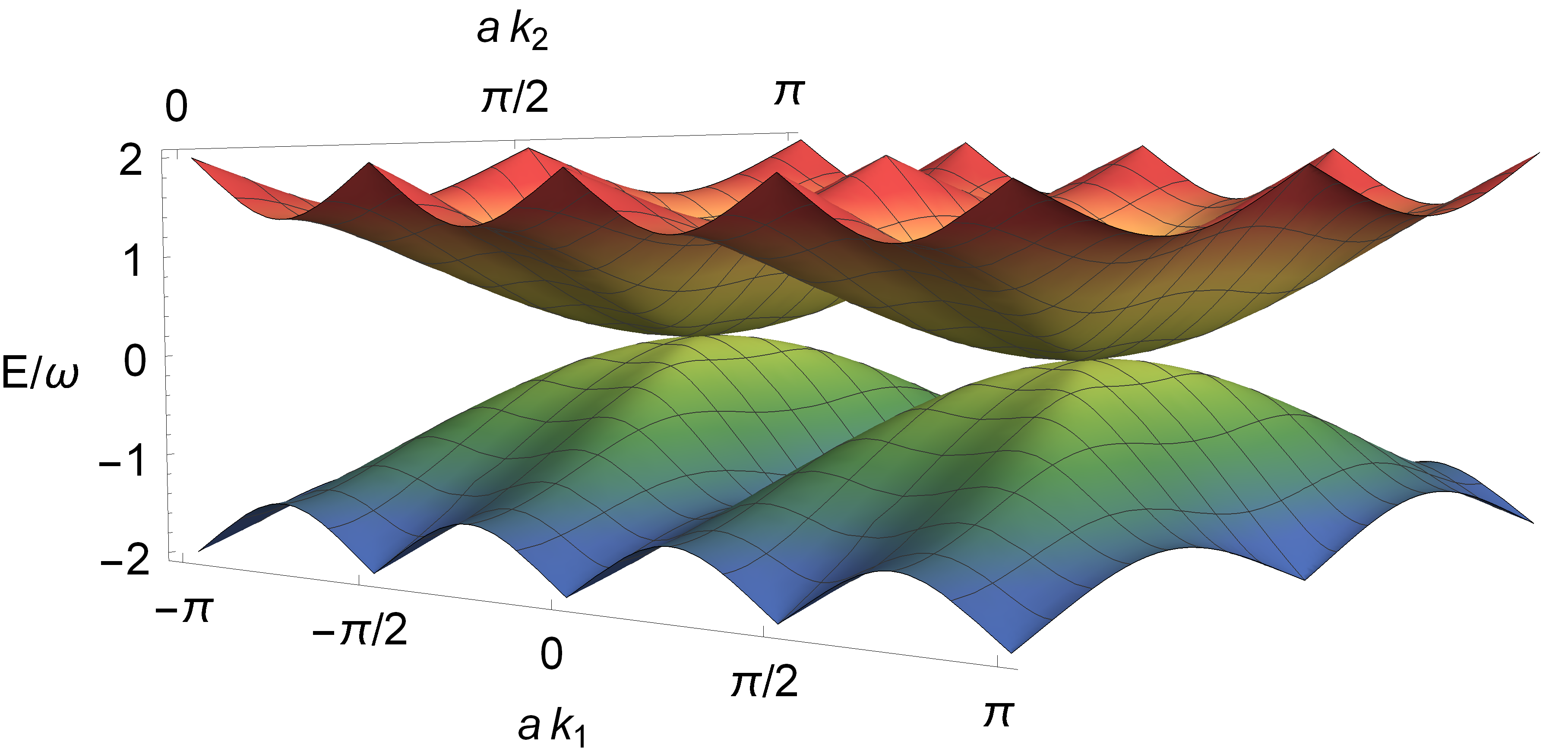}
	\llap{
  \parbox[b]{4cm}{$(b)$\\\rule{0ex}{3.9cm}
  }}
	\\
	\vspace{0.5cm}
  	\includegraphics[width=.48\textwidth]{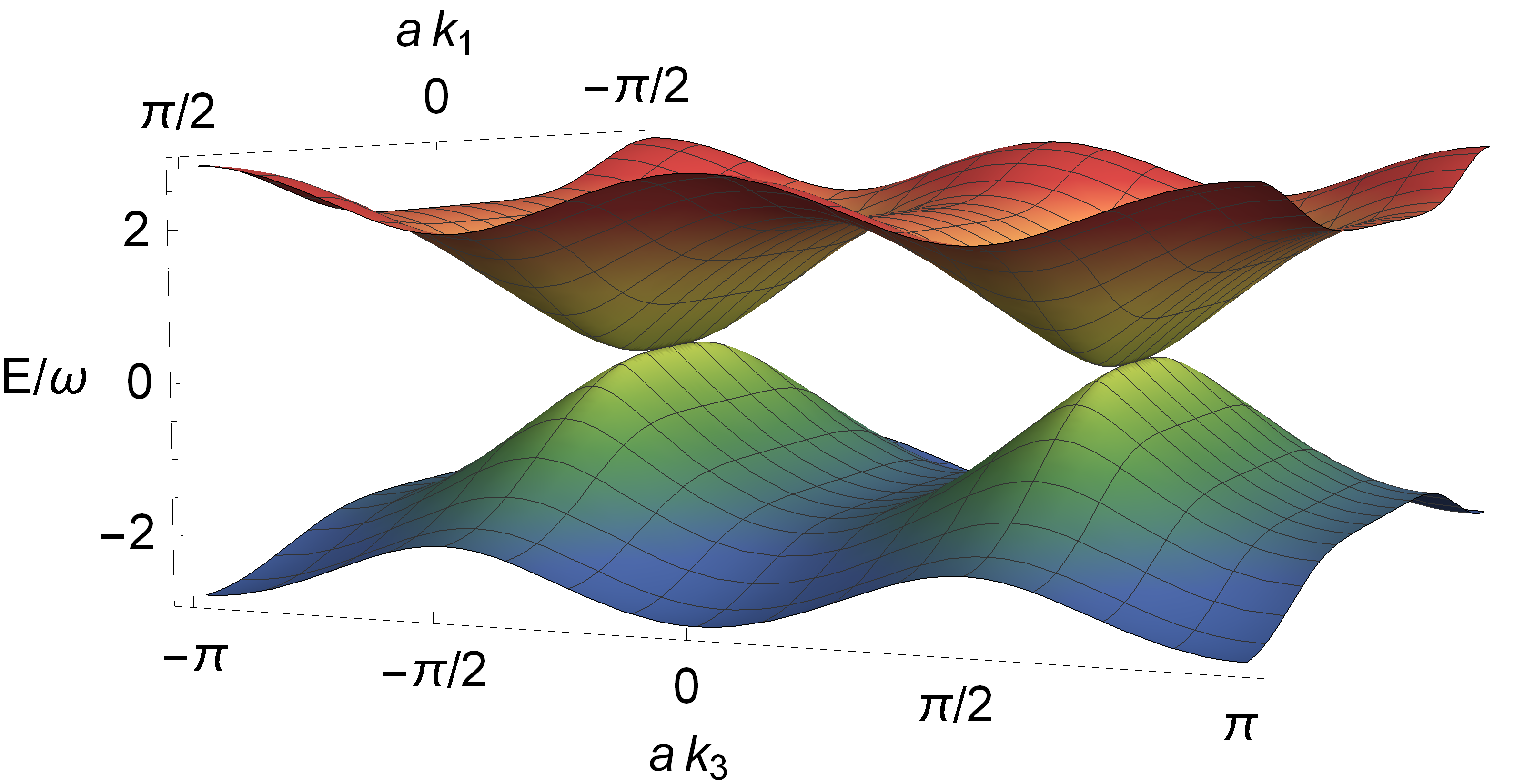}
	\llap{
  \parbox[b]{5cm}{$(c)$\\\rule{0ex}{3.9cm}
  }}\hfill
  	\includegraphics[width=.48\textwidth]{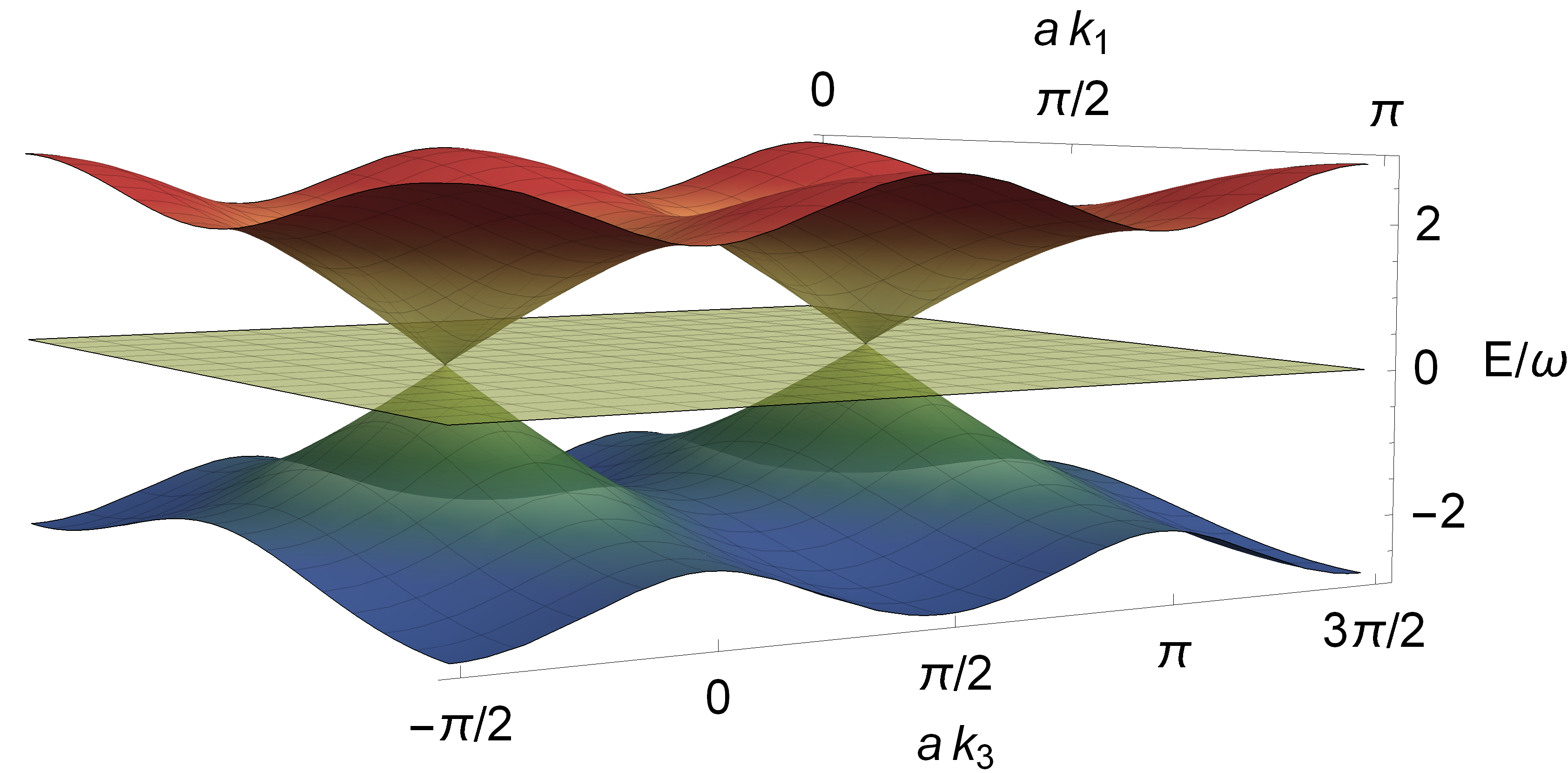}
	\llap{
  \parbox[b]{4cm}{$(d)$\\\rule{0ex}{3.9cm}
  }}
\caption{
Typical dispersions of tight-binding lattice models corresponding to the classes of topological semimetals considered in this paper. $\omega$ is an arbitrary energy scale and $\texttt{a}$ labels the lattice spacing. $(a)$ Dispersion as a function of $k_1$ and $k_2$ of a single-Weyl semimetal corresponding to the model in \cite{lepori2015}: {\color{black} it} displays pairs of linearly dispersing Weyl cones with opposite chirality in its Brillouin zone. $(b)$ Spectrum of a double-Weyl semimetal from the lattice model in \cite{lepori2016}: the energy is plotted as a function of $k_1$ and $k_2$, which are the directions with quadratic dispersion. The value of $k_3$, {\color{black} along which the dispersion is linear,} has been fixed at the band touching points. Around the double-Weyl points, the system displays a $C_4$ rotational symmetry {\color{black} with axis along the $k_3$ direction}. $(c)$ Typical spectrum of a triple-Weyl semimetal as a function of $k_1$ and $k_3$. The dispersion is cubic in $k_1$ (and $k_2$, not shown) and linear in $k_3$. $k_2$ has been chosen at the band touching point. $(d)$ Dispersion of the triple-point semimetal model in \cite{burrello17}. The triple point crossings are band touching points connecting a flat band at zero-energy and two linearly dispersing bands.
}
\label{fig:disp}
\end{figure}

Finally it is also possible to create systems where three bands connect together in a topological node with double monopole {\color{black} charge} \cite{bradlyn2016}. This is the case of the triple-point semimetals with a typical dispersion depicted in Fig. \ref{fig:disp}($d$). If time-reversal symmetry is preserved, the upper and lower bands display a linear dispersion, whereas the central band is flat  (it may display as well a quadratic dispersion along all directions, but the gradient of the energy with momentum vanishes in the triple-point crossing).  When time-reversal symmetry is broken, also the central band may acquire a dispersion \cite{burrello17}. 

{\color{black} In the following, our analysis of the  axial anomalies which characterize these models is based on lagrangian descriptions of these systems,  for energies close to the band-touching points}, where the lattice effect can be considered negligible.}

\section{Overview of the results}
\label{overview}

Our results are summarized in this section. In order to fix the notation, it is useful to add a brief  introductory note.

\subsection{Anomalous axial symmetry  for Weyl fermions} 

The concept of chiral anomaly developed originally in quantum field theory \cite{S,adler1969,bell1969}. Let us consider for instance the model of massless electrodynamics with lagrangian 
\beq
{\cal L} = {\overline \psi}(x) \left [ i \gamma^\mu D_\mu \right ] \psi (x) = 
  \psi^\dagger_R (x) \left [  i D_0 + i \, {\ner \sigma} \cdot {\bf D}  \right ] \psi_R (x) +  \psi^\dagger_L (x) \left [  i D_0 -  i \, {\ner \sigma} \cdot {\bf D}  \right ] \psi_L (x) \, ,
\label{2.1}
\eeq
where $D_\mu = \der_\mu + i A_\mu (x) $ represents  the covariant derivative,  $x \equiv x^{\mu} = (x^0,x^1,x^2,x^3)$, $\sigma^i$ (with $i=1,2,3$) denote the Pauli matrices,  and $\gamma^{\mu}$ (with $\mu=0,1,2,3$) the Dirac matrices, here in the Weyl representation. The lagrangian \eqref{2.1} is invariant under local vector $U(1)_V$ gauge transformations 
\be
e^{i \Lambda (x)} \in U(1)_V  \quad , \quad \left  \{  \begin{array}  {l@{ ~ } l}   
\psi_R(x) \longrightarrow \psi_R^\Lambda (x) = e^{i \Lambda (x)} \, \psi_R (x)  &     \\   \psi_L(x) \longrightarrow  \psi_L^\Lambda (x) = e^{i \Lambda (x) } \,  \psi_L (x)  
 &   \\
 A_\mu(x) \longrightarrow A^\Lambda_\mu (x) = A_\mu (x) - \der_\mu \Lambda (x)
\end{array} \right.  \; , 
\label{2.2}
\ee
and under global axial $U(1)_A$   transformations 
\be
e^{i \phi} \in U(1)_A  \quad , \quad \left  \{  \begin{array}  {l@{ ~ } l}   
\psi_R(x) \longrightarrow \psi_R^\phi (x) = e^{i \phi} \, \psi_R (x)  &     \\   \psi_L(x) \longrightarrow  \psi_L^\phi (x) = e^{-i \phi  } \,  \psi_L (x)  
 &   \\
 A_\mu(x) \longrightarrow  A_\mu (x) 
 \end{array} \right.  \; . 
\label{2.3}
\ee
At the classical level, the Noether axial  current $J^\mu_A $  corresponding to the symmetry (\ref{2.3}) satisfies  $\der_\mu J^\mu_A = 0$. However, at the quantum level the axial current is not conserved because of the so-called  axial anomaly. When the vector gauge symmetry  \eqref{2.2} is preserved, the axial anomaly  can be written in the form \cite{S,adler1969,bell1969} 
\be
\der_\mu J_A^\mu  (x)= - {1\over 4 \pi^2} \epsilon^{\mu \nu \tau \lambda} \der_\mu A_\nu (x) \der_\tau A_\lambda(x)  = - 
\frac{1}{4 \pi^2} \, {F^*}^{\tau \lambda} (x) \, F_{\tau  \lambda} (x)  \, ,
\label{2.4}
\ee
where $F_{\tau \lambda} (x)= \der_\tau A_\lambda (x)- \der_\lambda A_\tau(x) $ denotes the electromagnetic tensor, and ${F^*}^{\tau \lambda} (x) = \frac{1}{4} \epsilon^{\mu \nu  \tau  \lambda} \,  F_{\mu \nu} (x)$.  

 The same result is known {\color{black} to be also} valid for the single-Weyl semimetals \cite{nielsen1983,burkov2012}, where the theory \eqref{2.1} describes the low-energy dynamics around the nodal points; there the Pauli matrices act on  appropriate ``chiral indices'', generally labelling the sublattices that form the Weyl semimetal.

\subsection{Double-Weyl semimetals}
Double-Weyl semimetals  contain two inequivalent band touching points, protected by a symmetry $C_4$, with a linear dispersion relation along the direction connecting them, and a quadratic dispersion relation along the other two directions \cite{bernevig2012}. In real space-time, the corresponding  low energy lagrangian for the fermionic variables associated with these points takes the form   ($\hbar = c = 1$)  
\bea
{\cal L} &=&  \psi^\dagger_R (x) \left [  i D_0 - \alpha  \left ( D_1^2 - D_2^2 \right ) \sigma^1 - \mezzo \beta   \{ D_1 , D_2 \} \sigma^2  + i \gamma D_3 \sigma^3 \right ] \psi_R (x) \nonumber \\ 
&& + \, \psi^\dagger_L (x) \left [ i D_0 - \alpha  \left ( D_1^2 - D_2^2 \right ) \sigma^1 - \mezzo \beta  \{ D_1 , D_2 \} \sigma^2 -i \gamma D_3 \sigma^3 \right ] \psi_L (x) \; , 
\label{2.5}
\eea
where  $\alpha $, $\beta$ and $\gamma$ denote  three nonvanishing real constants (in the following, $\gamma$ and $\gamma^{\mu}$ should not be  confused). 
The  fermionic  (anticommuting) fields $\psi_R(x)$ and $\psi_L(x)$ have two components but, differently form the case of massless electrodynamics,  they do not represent Lorentz spinor fields. The variables  $\psi_R(x)$ and $\psi_L(x)$  are associated to opposite monopole charges, $\pm 2$, for the Berry flux around the corresponding nodes.   We call them ``right'' and ``left'' variables just for notational simplicity.    In terms of the four components field $\Psi (x) = (\psi_{R} (x), \psi_{L} (x))$, {\color{black} the lagrangian   \eqref{2.5} } can also be written as 
\beq
\mathcal{L}  = {\overline \Psi} (x)\, \left\lbrace i \gamma_0 \, D_0 -  \alpha \, \gamma_1 \gamma_5 \,  \left[D_1^2 - D_2^2  \right] - (\beta /2) \, \gamma_2 \gamma_5 \,  \{D_1 , D_2 \}  +i \gamma  \gamma_3 \, D_3   \right\rbrace \Psi (x) \, .
 \label{2.6}
\eeq
This expression makes it clear that, in this model, one has  a violation of the ``canonical''  form of the $PT$-symmetry, which is required to obtain stable multi-Weyl nodes \cite{wan11}-\cite{lepori2015}. 

The lagrangian  \eqref{2.5} is invariant under the local vector $U(1)_V$ gauge transformations shown in equation  \eqref{2.2},  and under global axial $U(1)_A$ transformations of equation \eqref{2.3}. The Noether axial current $J^\mu_A (x)$ corresponding to the symmetry (\ref{2.3}) can be written as 
\be
J^\mu_A (x) = J^\mu_R (x) - J^\mu_L (x) \; , 
\label{2.7} 
\ee
where 
 \be
\begin{tabular}{ccc} 
$J^0_R = \psi^\dagger_R  \psi_R$   &  & $J^0_L =    \psi^\dagger_L  \psi_L $ \\
$J_R^1 =  i  \psi^\dagger_R \left [ \alpha \sigma^1  {\overleftrightarrow D}_1 + \mezzo \beta  \sigma^2  {\overleftrightarrow D}_2 \right ] \psi_R  $ & ~~~~~ &
 $ J^1_L =
i  \psi^\dagger_L \left [ \alpha \sigma^1  {\overleftrightarrow D}_1 + \mezzo \beta  \sigma^2  {\overleftrightarrow D}_2 \right ] \psi_L $ \\
$ J_R^2 =  i   \psi^\dagger_R \left [ - \alpha \sigma^1  {\overleftrightarrow D}_2 + \mezzo \beta  \sigma^2  {\overleftrightarrow D}_1 \right ] \psi_R$ & &
$J^2_L = 
  i   \psi^\dagger_L \left [ - \alpha \sigma^1  {\overleftrightarrow D}_2 + \mezzo \beta  \sigma^2  {\overleftrightarrow D}_1 \right ] \psi_L $ \\ 
  $J^3_R =  \gamma \,  \psi^\dagger_R  \sigma^3   \psi_R$ & & $J^3_L  = -  \gamma \,  \psi^\dagger_L  \sigma^3   \psi_L $ \\ 
\end{tabular}
 \label{2.8}
\ee
in which  we have introduced the notation  $
\psi^\dagger \, {\overleftrightarrow D}_j \psi  \equiv 
 \psi^\dagger  \left ( D_j \psi \right )  - \left ( D_j  \psi^\dagger \right )  \psi $.  
 
 In this article we show that, for arbitrary nonvanishing values of $\alpha$, $\beta$ and $\gamma$,  when the gauge invariance  (\ref{2.2}) is maintained  the axial anomaly is given by 
\beq
 \partial_{\mu} J_A^{\mu} (x)=   -  \, 2 \, \Theta (\alpha , \beta , \gamma ) \frac{1}{4 \pi^2} \, \epsilon^{\mu \nu \tau \lambda} \der_\mu  A_\nu (x) \der_\tau A_\lambda (x) \; , 
 \label{2.9}
\eeq
where   
\be
\Theta (\alpha , \beta , \gamma )  = {\frac{\alpha \beta \gamma }{| \alpha \beta \gamma |}} \; . 
\label{2.10}
\ee
For fixed values of $\alpha$, $\beta$ and $\gamma$,  expression (\ref{2.9}) is twice the chiral anomaly \eqref{2.4} that one finds in  massless electrodynamics. In section~\ref{general} and section~\ref{perturb}, we  demonstrate  equation (\ref{2.9})  by means of a one-loop calculation based on Feynman diagrams.  Then the same result is  rederived by means of  the Nielsen-Ninomiya procedure  in section~\ref{NN-procedure},   and by means of the Atiyah-Singer  index argument in section~\ref{index}. 

The expression for the axial anomaly in equation  \eqref{2.9} makes  its  Lorentz independence  explicit. The Minkowski metric tensor $\eta_{\mu \nu}$ does not appear in equation \eqref{2.9} while the tensor $ \epsilon^{\mu \nu \tau \lambda}$ has  a "cohomological" origin ---as described in section \ref{general}---  in view of the fact that  the anomaly can be described by a differential  4-form.

\subsection{Triple-Weyl semimetals}

Triple-Weyl semimetals contain two inequivalent band touching points, protected by a symmetry $C_6$, with a linear dispersion relation along the direction connecting them and a cubic dispersion relation along the two remaining  directions \cite{bernevig2012}. The corresponding  low energy lagrangian can be written in real space-time as
\bea
{\cal L}  &=& \psi^\dagger_R (x) \left[  i D_0 - \alpha  \left(D_1^3 - {\mathcal S} \big[D_2^2  ,   D_1 \big] \right) \sigma^1 - \beta \left(D_2^3 - {\mathcal S} \big[D_1^2  ,   D_2 \big] \right) \sigma^2   + i \gamma D_3 \sigma^3 \right ] \psi_R (x)  + \nonumber \\ 
{}\nonumber\\
&&+ \, \psi^\dagger_L (x) \left[  i D_0 - \alpha  \left(D_1^3 - {\mathcal S} \big[D_2^2  ,   D_1 \big] \right) \sigma^1 - \beta \left(D_2^3 - {\mathcal S} \big[D_1^2  ,   D_2 \big] \right) \sigma^2   -  i \gamma D_3 \sigma^3 \right ] \psi_L (x) \; , 
\label{2.11}
\eea
where the symbol $\mathcal{S} [P^2,Q] \equiv PQP+PPQ+QPP$ implements the correct symmetrization of the covariant derivatives, and $\alpha $, $\beta$ and $\gamma$ denote   three nonvanishing real constants. In this case also $\psi_R(x)$ and $\psi_L(x)$ represent two-components anticommuting fields. 
 Precisely like  the case of the double-Weyl semimetals, the lagrangian \eqref{2.11} is invariant  under  local vector $U(1)_V$ gauge transformations  \eqref{2.2} and under  global axial $U(1)_A$ transformations  \eqref{2.3}. By means of  the Nielsen-Ninomiya procedure and the Atiyah-Singer  index argument,  in section~\ref{3-Weyl} we shown that, when the gauge invariance  (\ref{2.2}) is maintained,  the axial anomaly is given by 
\beq
 \partial_{\mu} J_A^{\mu} (x)=   -  3 \, \Theta (\alpha , \beta , \gamma ) \frac{1}{4 \pi^2} \, \epsilon^{\mu \nu \tau \lambda} \der_\mu  A_\nu (x) \der_\tau A_\lambda (x) \, .
 \label{2.12}
\eeq

The results in equations \eqref{2.4}, \eqref{2.9} and \eqref{2.12}
have been first inferred  in \cite{son2012}
 through a semiclassical calculation based on the kinetic theory of Landau Fermi liquids,  and in \cite{li2016} by means of  numerical and analytical approaches. In \cite{li2016}, the authors analyzed numerically the case with parallel electric and magnetic fields, and presented analytical calculations in the spirit of Nielsen and Ninomyia \cite{nielsen1983} for a system in which both fields are aligned along the direction of linear dispersion and in the presence of a full SO(2) rotational symmetry.
Finally, equation \eqref{2.9} has also been verified in \cite{shen2017} through a field theoretical approach, based on the Fujikawa's method (see for example \cite{Nakahara,fujikawa2017}), by evaluating the chiral anomaly for a double-Weyl point with SO(2) rotational symmetry. In all these papers, in the  expression  of the anomaly ---which has been derived in these articles---   
the $\Theta (\alpha , \beta , \gamma )$ factor \eqref{2.10} is missing. But the  absolute value   of the anomaly, which has been proposed in \cite{li2016,shen2017,son2012},  appears to be correct. The difference between the two derivations is that, in \cite{li2016,shen2017,son2012}, the left field has been considered set, since the beginning, by the condition $\alpha \beta \gamma > 0$.

The derivations in \cite{li2016,shen2017} do not fully explain the quantization of the anomaly, with  topological charge $\mathcal{N}\Theta (\alpha , \beta , \gamma )$, and, from a more fundamental point of view, do not clarify why the anomaly for multi-Weyl semimetals is proportional to the differential form $F (x) \wedge F (x)$, in spite of the breaking of Lorentz covariance of the corresponding low energy lagrangians.  In facts, doubts \cite{shizuya87,joglekar87} have been cast upon the use of the  regularization scheme of the path-integral measure exploited in the Fujikawa's method \cite{shen2017} in cases different from the standard Weyl theory.

\subsection{Triple-point semimetals}

Triple-point semimetals  \cite{bradlyn2016} are characterized by two zero-energy points in which three bands join. The lagrangian of the low energy model takes the form 
 \bea
{\cal L} &=&  \psi^\dagger_R (x)\,  i\left [   D_0 - v   M_1 D_1 - v  M_2  D_2   - v M_3 D_3  \right ] \psi_R (x) \nonumber \\ 
&& + \, \psi^\dagger_L (x)\,  i \left [ 
  D_0 - v  M_1 D_1 - v  M_2  D_2   + v M_3 D_3  \right ] \psi_L (x) \; , 
\label{2.13}
\eea
 where the  real parameter  $v$ is positive and $D_\mu = \der_\mu + i A_\mu (x)$. The fermionic fields  $\psi_R(x)$ and $\psi_L(x)$ have three components.   The three matrices $M_j$ are given by 
 \be
 M_1 = M_1(\theta) = \begin{pmatrix}  0 & 0 & 0 \\ 0 & 0 & i e^{i \theta} \\ 0 & -i e^{-i\theta} & 0 
 \end{pmatrix} \quad , \quad 
 M_2 = M_2(\theta) = \begin{pmatrix}  0 & 0 & i e^{-i\theta} \\ 0 & 0 & 0 \\ -i e^{i \theta} & 0 &  0 
 \end{pmatrix} \quad , \quad 
 M_3 = M_3(\theta) = \begin{pmatrix}  0 & i e^{i\theta} & 0 \\  - i e^{- i \theta} & 0 & 0 \\ 0 & 0 & 0 
 \end{pmatrix} \; ,  
 \label{2.14}
 \ee
in which the angle $\theta $  is a parameter of the model which breaks time-reversal symmetry. The $M_j$ matrices can be interpreted ad deformed generators of the rotation group in the adjoint representation; in our notation these matrices 
satisfy the commutation relations:
\be
\big[M_j(\theta_1),M_k(\theta_2)\big] = i \epsilon_{jk\ell }  \, M_\ell (-\theta_1- \theta_2) \, .
\eeq

 The lagrangian \eqref{2.13} is invariant under local vector gauge transformations \eqref{2.2} and under global axial transformations \eqref{2.3}. When the vector gauge invariance is preserved, 
the axial anomaly is found to be 
\beq
 \partial_{\mu} J_A^{\mu} (x)=    \frac{\cos (3 \theta)}{| \cos (3 \theta) | } \,  \frac{1}{4 \pi^2} \, \epsilon^{\mu \nu \tau \lambda} \der_\mu  A_\nu (x) \der_\tau A_\lambda (x) \; .  
 \label{2.15}
\eeq
This expression is derived in section~\ref{3-point} by means of the Nielsen-Ninomiya and Atiyah-Singer methods.

\section{Gauge anomalies in perturbative quantum field theory}
\label{general}

Before proceeding with the direct computation of the anomaly, it is useful to discuss the relationship between the so-called  left-right and  vector-axial possible forms of the anomaly, together with a few general properties of gauge anomalies.  

\subsection{Perturbative approach}

In order to simplify the exposition   and avoid repetitions, in the following discussion we concentrate directly   on the double-Weyl model  \eqref{2.5}; but  the results of this section clearly have a general validity.   
It is convenient to examine first the lagrangian terms for the massless fermionic fields $\psi_R(x)$ and $\psi_L(x)$  separately; afterwards, the anomalous behaviours of their corresponding  one-loop diagrams will be  combined in order to determine the desired axial anomaly. Let us  consider  the ``right-handed''  component $\psi_R(x)$. 
In order to simplify the exposition, in the intermediate steps of the computation  the gauge field coupled with $\psi_R(x)$ will be denoted by $V_\mu (x)$.  We  shall recover the previous $A_{\mu} (x)$ notation  at the end of the present section.  The corresponding lagrangian density ${\cal L}_R$ takes the  form 
\bea
{\cal L}_R &=& \psi^\dagger_R (x) \, \Pi_R (V) \,  \psi_R (x) \nonumber \\ 
&=&  \psi^\dagger_R (x) \left [  i D_0 - \alpha  \left ( D_1^2 - D_2^2 \right ) \sigma^1 - \mezzo \beta   \{ D_1 , D_2 \} \sigma^2  + i \gamma D_3 \sigma^3 \right ] \psi_R (x) \; , 
\label{3.1}
\eea
where $D_\mu = \der_\mu + i V_\mu (x)$. The function ${\cal L}_R$ is invariant under local $U(1)_R$ gauge transformations 
\be
 \hbox{local } U(1)_R  : \quad \left  \{  \begin{array}  {l@{ ~ } l}   
\psi_R(x) \longrightarrow  e^{i \theta_R (x)} \, \psi_R (x)  &    \\  ~ & ~ \\  
 V_\mu (x) \longrightarrow  V_\mu (x) - \der_\mu \theta_R (x) \; . 
\end{array} \right.  
\label{3.2}
\ee
 The operator $\Pi_R (V)$ which enters equation (\ref{3.1}) can be written as the sum of two terms, $\Pi_R (V) = \Pi_R(0) + \widetilde \Pi_R (V)$,  in which the free part $ \Pi_R(0) = i \der_0 - \alpha  \left ( \der_1^2 - \der_2^2 \right ) \sigma^1 - \beta  \der_1  \der_2 \,  \sigma^2  + i \gamma \der_3 \sigma^3$ does not depend on $V_\mu$. Therefore the free spinor propagator \cite{IZ,weinberg,peskin} is given by  
 \be
 \WT{\psi_R (x)\> \psi}\null_{\! \! R}^{\! \!  \dagger}  (y)  \equiv 
 \bra{0} \hbox{T }\psi_R(x) \psi^\dagger_R (y) \ket{0}  = i \! \int \! \frac{d^4p}{(2 \pi)^4} e^{-i p (x-y) } \, \frac{p_0 + \alpha (p_2^2 - p_1^2)\sigma^1 - \beta p_1 p_2 \sigma^2 - \gamma p_3 \sigma^3}{p_0^2 - \alpha^2 (p_1^2 -p_2^2)^2 -\beta^2 p_1^2p_2^2 - \gamma^2 p_3^2 + i \varepsilon }
  \; , 
\label{3.3}
\ee
 and the interaction component of the action takes the form  
 \bea
S_{RI} [V] &=& \int d^4x \; \psi^\dagger_R (x) \,  \widetilde \Pi_R (V) \, \psi_R (x) \nonumber \\ 
&=& \int d^4x \; \psi^\dagger_R (x) \Bigl \{  - V_0 + \alpha   \left [  V_1^2 - V_2^2 +i (\der_2 V_2) + 2 i V_2 \der_2 -i (\der_1V_1) -2 i V_1 \der_1   \right ] \sigma^1 \nonumber \\
&& + \mezzo \beta  \left [ 2 V_1 V_2 -i (\der_1 V_2) -2i V_2 \der_1  -i (\der_2 V_1) -2 i V_1 \der_2  \right ] \sigma^2  - \gamma V_3 \sigma^3 \Bigr \} \psi_R (x) \; . 
\label{3.4}
 \eea
 In expression \eqref{3.3}, the Feynman $\varepsilon$-prescription \cite{IZ,peskin} has been introduced in order to guarantee causality and energy positivity.   Let $i \Gamma_R [V]$ denote the sum of the connected one-loop vacuum-to-vacuum diagrams \cite{IZ,weinberg,peskin} of   the $\psi_R (x)$ field in the presence of the classical background field $V_\mu (x)$, 
\be
e^{i \Gamma_R [V]} = \langle 0 | \hbox{ T } e^{i S_{RI}[V]} \, | 0 \rangle 
\; , 
\label{3.5}
\ee
where the symbol T denotes the Wick time-ordering. From the definition (\ref{3.5}) it follows that, under a gauge transformation $V_\mu (x) \rightarrow V_\mu(x) - \der_\mu \theta_R(x)$, the infinitesimal variation $\delta_{\theta_R} \Gamma_R [V]$ of $\Gamma_R [V]$ is given by the sum of the connected diagrams 
\be
\delta_{\theta_R} \Gamma_R[V] =  \int d^4 x  \,  \der_\mu \theta_R (x) \langle0 | \hbox{ T }   J_R^\mu (x)  e^{i S_{RI}} \, | 0 \rangle^c = - \int d^4 x  \,   \theta_R (x) \, \langle \der_\mu J^\mu_R (x)\rangle \; .  
\label{3.6}
\ee
 The gauge invariance of the lagrangian ${\cal L}_R$  under the transformations in \eqref{3.2} would suggest that $\Gamma_R[V]$ also is gauge invariant, and consequently $\delta_{\theta_R} \Gamma_R[V] = 0$.   However, because of ultraviolet divergences, the functional  $\Gamma_R[V]$ is not well defined. Therefore the central question   is whether one can define or not a renormalized  $\Gamma_R[V]$ which is gauge invariant.  If a  renormalized gauge invariant $\Gamma_R[V]$ exists, then the gauge symmetry (\ref{3.2}) is not anomalous and $\der_\mu J^\mu_R =0$. Otherwise,  $\delta_{\theta_R} \Gamma_R[V] \not= 0 $, and  one finds an anomaly  
 \be
 \der_\mu J^\mu_R (x) =  \langle \der_\mu J_R^\mu (x) \, \rangle = {\cal P}_R (V) \not= 0 \; ,  
 \label{3.7}
 \ee
 where, in agreement with the action principle, ${\cal P}_R (V)$ is a local polynomial  of the field $V_\mu (x)$ and of its space-time derivatives.   Usually, the construction of a renormalized  $\Gamma_R[V]$ consists of two steps: 
\begin{enumerate}
\item definition of a regularized  functional $ \Gamma^{reg}_R[V]$, which depends on   a  cutoff, 
\item introduction of local counterterms $\Gamma_{ct}[V]$, containing in general both divergent and finite parts, which reabsorb the ultraviolet divergences. 
\end{enumerate}
The renormalized  functional $\Gamma_R[V]$, which is well defined (free of divergences),  corresponds to the sum $ \Gamma^{reg}_R[V] + \Gamma_{ct}[V]$ in the limit in which the cutoff is removed. The particular choice of the regularisation is totally irrelevant. In the renormalization procedure, the freedom of adding finite local counterterms  completely removes  the    dependence of the result on the particular choice of the regularization, because two different regularizations differ (in the limit of removed cutoff) by the sum of finite local   counterterms \cite{IZ,weinberg,peskin}.   Thus the expression of the polynomial  ${\cal P}_R(V)$  of $V_\mu (x)$, which appears in 
\be
\delta_{\theta_R} \Gamma_R[V] = - \int d^4 x  \,  \theta_R (x) {\cal P}_R (V)  \; ,  
\label{3.8}
\ee
is not uniquely determined,  because one can  add  the gauge variation $\delta_{\theta_R} L_{ct} [V]$ of some finite local counterterm $L_{ct}[V]$ to the integral $\int \theta_R {\cal P}_R$.  Consequently,  if  $\int \theta_R {\cal P}_R$ can be written as the gauge variation of a local counterterm, then there is no anomaly since, by introducing the appropriate counterterm,  one can define a renormalized  gauge invariant functional $\Gamma_R[V]$. The existence of the anomaly means that  expression \eqref{3.8} cannot be written as the gauge variation of a local term. In this case,   even if one can modify the expression of $\delta_{\theta_R} \Gamma_R[V]$ ---by means of local counterterms---  one cannot eliminate ${\cal P}_R(V)$.  Precisely for this reason, the existence of the anomaly does not depend on the choice of the regularization. 
 Of course, the presence or the absence of the  anomaly is  determined by  the lagrangian \eqref{3.1} which specifies how the field $\psi_R (x)$ interacts with the gauge field.     

\subsection{Anomaly as a solution of a cohomological problem}

By developing the consequences of the Wess-Zumino consistency conditions \cite{WZ}, it has been found \cite{BZ,MST,DB,BL,BPT,SO} that the search of possible nontrivial solutions to equation (\ref{3.8}) can actually be reduced  to a cohomological problem. Indeed, the gauge variation of any function  $ f[V_\mu] $ can be represented by the action on $ f[V_\mu] $ of a nilpotent BRST \cite{BRST} operator ${\cal T}$, which is defined (in the abelian case) by the relations  ${\cal T}\, V_\mu (x) = - \der_\mu c(x)$ and ${\cal T} \, c (x) = 0 $, in which the anticommuting variable $c(x)$ takes the place of the gauge parameter $\theta_R(x)$.   Since the anomaly is determined by the gauge variation of $\Gamma_R [V]$, the anomaly is $\cal T$-closed, but it is not   $\cal T$-exact in the set of local counterterms $\{ L_{ct} [V] \}$.  The anomaly then represents a nontrivial solution of the following cohomological problem 
\be
{\cal T} \left ( \int \, c (x) {\cal P}_R (V) \right ) = 0 \quad , \quad \int c(x) {\cal P}_R (V) \not= {\cal T} L_{ct} [V]\quad , \quad \hbox{with} \quad  {\cal T}^2 = 0 \; . 
\label{3.9}
\ee
In this general approach, the gauge fields are described by differential forms, $V = V_\mu (x) dx^\mu $; no Lorentz invariance is assumed  and only the properties of the gauge transformations group  enter  the solutions. In this way,  the possible forms of the gauge anomalies can generally be determined without the need of introducing any  corresponding field theory model.  More precisely, all the local polynomials of the field $V_\mu (x)$,  which are not equal to the gauge variation of a local counterterm,  have been produced. The only parameter which is not  fixed  \`a  priori by cohomological arguments is the overall normalization factor  of each polynomial. The value of this normalization factor is specified by the lagrangian of each particular model. 

In the case of the abelian gauge symmetry $V_\mu (x) \rightarrow  V_\mu (x) - \der_\mu \theta_R (x)$, equation (\ref{3.8}) can always \cite{S,adler1969,bell1969,ABA,BA,ZWZ,BAZ} be written   in the form  
 \be
\delta_{\theta_R} \Gamma_R[V] =  - {\cal N} {1\over 24 \pi^2} \int d^4 x  \, \epsilon^{\mu \nu \tau \lambda}   \der_\mu \theta_R (x)  V_\nu (x) \der_\tau V_\lambda (x)  \; ,  
\label{3.10}
\ee
where $\cal N$ represents an overall  multiplicative factor that must be computed.  
 If ${\cal N} =0$, there is no anomaly. One can easily verify that, when ${\cal N} \not= 0$, expression \eqref{3.10} cannot be written as the gauge variation of a local counterterm. Therefore the anomaly exists for ${\cal N} \not= 0$.  In general, the coefficient $\cal N$ takes integer values; this point will be discussed in section~\ref{quantization}. For instance, in the case of a relativistic right-handed Weyl spinor minimally coupled  with the gauge field $V_\mu(x)$, one finds ${\cal N} =1$.   
 
 In the case analyzed in this section, the value of $\cal N$   is determined by the specific form of the  lagrangian density \eqref{3.1}; in particular, the anomaly is specified by the structure of the operator $\Pi_R (V)$. By a direct computation, we will show  that ${\cal N} \not= 0 $. Even if in our case the field $\psi_R(x)$ does not represent a spinor field, in agreement with the standard notation,   expression \eqref{3.10} will be called the chiral anomaly. 
 
 \subsection{Axial anomaly}
 
 In order to derive  the general form of the $U(1)_A $ axial anomaly in the double-Weyl model with  lagrangian density \eqref{2.5},   let us now consider  the field $\psi_L (x)$ and let us denote by $W_\mu (x) $ the gauge field which is coupled with $\psi_L (x)$ according to the lagrangian 
 \bea
 {\cal L}_L &=& \psi^\dagger_L (x) \, \Pi_L (W) \,  \psi_L (x) \nonumber \\
 &=&  \psi^\dagger_L (x) \left [  i D_0 - \alpha  \left ( D_1^2 - D_2^2 \right ) \sigma^1 - \mezzo \beta   \{ D_1 , D_2 \} \sigma^2  - i \gamma D_3 \sigma^3 \right ] \psi_L (x) \; , 
\label{3.11}
\eea
where $D_\mu = \der_\mu + i W_\mu (x)$. ${\cal L}_L$ is invariant under local $U(1)_L$ gauge transformations 
\be
 \hbox{local } U(1)_L  : \quad \left  \{  \begin{array}  {l@{ ~ } l}   
\psi_L(x) \longrightarrow  e^{i \theta_L (x)} \, \psi_L (x)  &    \\  ~ & ~ \\  
 W_\mu (x) \longrightarrow  W_\mu (x) - \der_\mu \theta_L (x) \; . 
\end{array} \right.  
\label{3.12}
\ee
Let $i \Gamma_L[W]$ be  the sum of the connected one-loop vacuum-to-vacuum diagrams of $\psi_L(x)$ in the presence of the classical field $W_\mu$. The infinitesimal variation $\delta_{\theta_L} \Gamma_L [W]$ of $ \Gamma_L[W]$ under the transformation 
 $W_\mu (x) \rightarrow W_\mu(x) - \der_\mu \theta_L (x)$ is strictly related with $\delta_{\theta_R} \Gamma_R [V]$.   Indeed, the lagrangian  for $ \psi_L(x)$ can be obtained from the lagrangian for  $\psi_R(x)$ by means of the substitution   $\sigma_3 \rightarrow - \sigma_3$. This means that the expression of the anomaly for $\psi_L(x)$ can be obtained from expression (\ref{3.10}) provided we introduce, in addition to the obvious change  of variables,   a change of the sign of the $x^3$-derivative, $\der_3 \rightarrow  - \der_3 $, and a change of the sign of the third component of the field $W_\mu(x)$, $W_3 (x) \rightarrow - W_3(x)$.   Therefore 
 \be
\delta_{\theta_L} \Gamma_L[W] =   {\cal N} {1\over 24 \pi^2} \int d^4 x  \, \epsilon^{\mu \nu \tau \lambda}   \der_\mu \theta_L (x)  W_\nu (x) \der_\tau W_\lambda (x)  \; .  
\label{3.13}
\ee
 
 Let us now consider the complete model with lagrangian \eqref{2.5} in which both $\psi_R(x)$ and $\psi_L(x)$ are present. 
 The gauge fields $V_\mu (x)$ and $W_\mu (x)$, that refer to the components of the group $U(1)_R \times U(1)_L$, can be written as combinations of the vector fields associated with the components of $U(1)_V \times U(1)_A$:
 \bea
 U(1)_V  \quad &,& \quad A_\mu (x) = {1\over 2} \left [ V_\mu (x)+ W_\mu (x)\right ] \nonumber \\ 
 U(1)_A  \quad &,& \quad B_\mu (x) = {1\over 2} \left [ V_\mu  (x)- W_\mu (x) \right ] \; . 
 \label{3.14}
 \eea
 The infinitesimal variation of  $\widetilde \Gamma [A,B] = \Gamma_R [V(A,B)] + \Gamma_L [W(A,B)]$ under the vector $U(1)_V$ transformation  $A_\mu (x) \rightarrow A_\mu (x)- \der_\mu \theta_V(x)$ is obtained by combining equations (\ref{3.10}) and (\ref{3.13})
 \be
 \delta_{\theta_V} \widetilde \Gamma [A,B] = - {\cal N} {1\over 6 \pi^2} \int d^4 x  \, \epsilon^{\mu \nu \tau \lambda}   \der_\mu \theta_V (x)  B_\nu (x) \der_\tau A_\lambda (x)  \; ,   
\label{3.15}
 \ee
 while the infinitesimal variation of $\widetilde \Gamma [A,B]$ under the $U(1)_A$ axial transformation    $B_\mu (x) \rightarrow B_\mu (x) - \der_\mu \theta_A (x)$ turns out to be  
  \be
  \delta_{\theta_A} \widetilde \Gamma [A,B] = - {\cal N} {1\over 12 \pi^2} \int d^4 x  \, \epsilon^{\mu \nu \tau \lambda}   \der_\mu \theta_A (x) \left [  A_\nu (x) \der_\tau A_\lambda (x) + B_\nu (x) \der_\tau B_\lambda (x)  \right ]  \; . 
  \label{3.16}
  \ee
  Let us introduce the  functional 
 \be 
 \Gamma [A,B] = \widetilde \Gamma [A,B] + L[A,B] \, 
 \label{3.17}
 \ee
 where the finite  local counterterm $L [A,B]$ is given by 
 \be
 L[A,B] = - {\cal N} {1\over 6 \pi^2} \int d^4 x  \, \epsilon^{\mu \nu \tau \lambda}  A_\mu (x) B_\nu (x) \der_\tau A_\lambda (x) \; . 
 \label{3.18}
 \ee
 The infinitesimal variations of $\Gamma [A,B]$ under $U(1)_V \times U(1)_A$ transformations take the form    
 \be
 \delta_{\theta_V} \Gamma [A,B] = 0 \; , 
 \label{3.19}
 \ee
and 
{ \color{black}
\be
\delta_{\theta_A} \Gamma [A,B] =  {\cal N} {1\over 12 \pi^2 } \int d^4 x  \, \epsilon^{\mu \nu \tau \lambda}    \theta_A (x) \left [  \der_\mu B_\nu (x) \der_\tau B_\lambda (x) + 3 \der_\mu A_\nu (x) \der_\tau A_\lambda (x)  \right ] \; . 
\label{3.20}
\ee 
}
One can easily verify that expression \eqref{3.20} is not the gauge variation of a local counterterm.  Equation (\ref{3.19}) shows that the subgroup $U(1)_V$ is anomaly  free; consequently,  the vector gauge invariance \eqref{2.2} is preserved  and the corresponding local gauge theory is  consistent.   The gauge  anomaly only concerns the axial subgroup $U(1)_A$.  In the model which is described by the lagrangian density (\ref{2.5}), the field $B_\mu (x)$  is vanishing; therefore expression (\ref{3.20}) evaluated at $B_\mu (x)=0 $  gives
\be
\delta_{\theta_A} \Gamma [A,B] \Bigr |_{B_\mu = 0} = {\cal N} {1\over 4 \pi^2 } \int d^4 x  \, \epsilon^{\mu \nu \tau \lambda}    \theta_A (x) \der_\mu  A_\nu (x) \der_\tau A_\lambda (x)  \; . 
\label{3.21}
\ee
So, in the double-Weyl model \eqref{2.5}, the  divergence of the axial current  ---or the expression of the axial anomaly--- takes the form    
\be
\der_\mu J^\mu_A (x) = - {\cal N} {1\over 4 \pi^2} \epsilon^{\mu \nu \tau \lambda} \der_\mu  A_\nu (x) \der_\tau A_\lambda (x)  \; ,  
\label{3.22}
\ee
which is in agreement with equation (\ref{2.9});  the value of $\cal N$ remains to be computed.

Equation (\ref{3.22})  shows that, in multi-Weyl semimetals as well as in other generic field theory models,  the axial anomaly ---if present---  is proportional to the standard axial anomaly of massless electrodynamics.  Indeed,  on the one hand,  the cohomological problem \eqref{3.9} admits a universal nontrivial  solution and, on the other hand, in the presence of $U(1)_R \times U(1)_R$ symmetry the vector  $U(1)_V$ invariance  is required to be preserved.   Thus  no functional modification ---due to the absence of effective Lorentz covariance of the classical lagrangian---  appears in the axial anomaly.  

We mention finally that if an explicit gauge symmetry breaking is induced, modifications from the expression in \eqref{3.22} are expected;  one example has been considered recently in \cite{brien2017}.

\section{Perturbative computation}
\label{perturb}

In this section we shall derive the expression \eqref{3.10} of the chiral anomaly for the double-Weyl model  by means of perturbation theory. In particular, the value  of $\cal N$ ---appearing in equation \eqref{3.10}--- will be determined. This means that, as it has been shown in section~\ref{general}, the result of this section  provides a proof of equation \eqref{2.9}. 
 
 \subsection{Regularization}

 As it has been shown in  section~\ref{general}, the origin of the chiral anomaly is represented  by the nontriviality of the gauge variation \eqref{3.10} of the functional $\Gamma_R [V]$. 
The sum of the connected one-loop diagrams  entering the definition \eqref{3.5} is given by \cite{IZ,weinberg,peskin}

\bea
 i \Gamma_R[V] &=& - \sum_{n=1}^\infty \, \frac{1}{n}\, \hbox{Tr} \left [  i    \WT{\psi_R \> \psi}\null_{\! \! R}^{\! \!  \dagger} \, \widetilde \Pi_R (V)       \right ]^n    \nonumber \\
 &=& \, - \sum_{n=1}^\infty \frac{1}{n} \, \hbox{Tr}\left[\int d^4 x_1 \ldots d^4 x_{n} \, \,  \bra{x_1} i    \WT{\psi_R \> \psi}\null_{\! \! R}^{\! \!  \dagger} \, \widetilde \Pi_R (V) \ket{x_2}  \cdots \bra{x_{n}} i    \WT{\psi_R \> \psi}\null_{\! \! R}^{\! \!  \dagger} \, \widetilde \Pi_R (V) \ket{x_1} \right ]  \; , 
 \label{4.1}
 \eea
 where, in agreement with the Schwinger notations  \cite{S},  the symbol  Tr represents   the trace  
\be
\hbox{Tr } (Q) = \int d^4 x \, \hbox{tr}\, \langle x | \, Q \,  | x \rangle  \; , 
\label{4.2}
\ee
in which  Tr denotes the trace over the indices of the sigma matrices. Since the fermion propagator takes the form shown in equation \eqref{3.3}, equality \eqref{4.1} can be written as 
\be
 i \Gamma_R[V] = \hbox{Tr} \ln \left \{ 1 + 
 \frac{i \der_0 + \alpha  \left ( \der_1^2 - \der_2^2 \right ) \sigma^1 + \beta  \der_1  \der_2 \,  \sigma^2  - i \gamma \der_3 \sigma^3}{- \der_0^2 - \alpha^2 (\der_1^2 -\der_2^2)^2 -\beta^2 \der_1^2\der_2^2 + \gamma^2 \der_3^2 + i \varepsilon } \widetilde \Pi_R (V)
 \right \} \; . 
 \label{4.3}
\ee
Indeed the expansion of expression \eqref{4.3} in powers of $\widetilde \Pi_R(V)$ coincides with equation \eqref{4.1}. 
The terms of the sum \eqref{4.1} which  correspond to the divergent Feynman diagrams  are not well defined. So we now introduce a regularisation. Let us recall that, if $y$ is a positive number, one has 
\be
\ln y + \hbox{ constant } = - \lim_{\epsilon \rightarrow 0} \int_\epsilon^\infty {ds \over s} e^{- s y} \; . 
\label{4.4}
\ee
 Therefore, according to  the Schwinger proper-time regularisation \cite{S}, the regularised one-loop functional is defined as  
\be
\Gamma^{reg}_R[V] = i  \int_\epsilon^\infty {d s \over s} \, \hbox{Tr} \left [ \, e^{s \Sigma  \Pi_R(V)} \,  \right ] + \hbox{ constant }  \; ,  
\label{4.5}
\ee
where the constant does not depend on $V_\mu $, $\Pi_R(V)$ is shown in equation (\ref{3.1}), and  the $\Sigma $ operator,  
\be
\Sigma = i \der_0 + \alpha  \left ( \der_1^2 - \der_2^2 \right ) \sigma^1 + \beta  \der_1  \der_2 \,  \sigma^2  - i \gamma \der_3 \sigma^3 \, , 
\label{4.6}
\ee
enters the definition of the propagator \eqref{3.3}.   The sign in the exponent of equation \eqref{4.5} is fixed by the positivity of the analytic extension of $- \Sigma \Pi_R (V) $ in the euclidean region for the momenta.   The parameter $\epsilon > 0 $ represents the cut-off, and the limit of vanishing cut-off is obtained by taking the    $ \epsilon \rightarrow 0 $ limit. 

\subsection{Gauge variation}

Under a gauge transformation $V_\mu (x) \rightarrow V_\mu (x) - \der_\mu \theta_R(x) $,  the infinitesimal variation of $\Gamma^{reg}_R[V]$ is given by 
\bea
\delta_{\theta_R} \Gamma^{reg}_R [V]&=&   i\,  \int_\epsilon^\infty {d s \over s} ( is) \hbox{Tr} \left \{ e^{s \Sigma \Pi_R[V]} \Sigma [  \theta_R ,  \Pi_R (V)] \right \} \nonumber \\
&=&  \int_\epsilon^\infty d s \, \hbox{Tr}\, 
\Bigl \{ e^{ s \Sigma \Pi_R  }  \Sigma \Pi_R \theta_R - e^{ s \Pi_R \Sigma }   \Pi_R \Sigma  \theta_R    \Big \} = \hbox{Tr}\, 
\Bigl \{ \left [ e^{ \epsilon \Sigma \Pi_R [V] } - 
e^{\epsilon \Pi_R[V]  \Sigma}  \right ] \theta_R \Bigr \}    \; . 
\label{4.7}
\eea
By means of the relation 
\bea
e^{X + Y} &=& e^X + \int_0^1 du \; e^{(1-u) X }
 Y e^{u X}+ \int_0^1 u du \int_0^1 dv \; e^{(1-u) X } Y e^{u(1-v) X } Y e^{uv X }\nonumber \\
  &&+ \int_0^1 u^2 du \int_0^1 vdv \int_0^1 dt \; e^{(1-u) X } Y e^{u(1-v) X } Y e^{uv (1 - t)X } Y e^{uvt X } + \cdots 
  \label{4.8}
 \eea
one  obtains 
\bea
\delta_{\theta_R} \Gamma^{reg}_R [V]&=&  \hbox{Tr}\,  \Bigl \{  \epsilon \int_0^1 du \; e^{(1-u) \epsilon \Pi^2 } \widetilde \Pi_{R} [V] e^{u \epsilon \Pi^2} [\theta_R , \Sigma] \nonumber \\ 
 &&+ \epsilon^2 \int_0^1 u du \int_0^1 dv \; e^{(1-u)\epsilon \Pi^2 } \widetilde \Pi_{R} [V] e^{u(1-v) \epsilon \Pi^2} \Sigma \widetilde \Pi_{R} [V] e^{uv \epsilon \Pi^2 } [\theta_R , \Sigma ]\nonumber \\
 &&+ \epsilon^3  \int_0^1 u^2 du \int_0^1 vdv \int_0^1 dt \; e^{(1-u) \epsilon \Pi^2 } \widetilde \Pi_{R} [ V] e^{u(1-v) \epsilon \Pi^2 } \Sigma \widetilde \Pi_{R} [V] \, \times \nonumber \\ 
&& \quad \times  e^{uv (1 - t)\epsilon \Pi^2 } \Sigma \widetilde \Pi_{R} [V] e^{uvt \epsilon \Pi^2 } [\theta_R , \Sigma ] + \cdots \Bigr \} \; ,  
\label{4.9}
\eea   
where 
\be
\Pi^2 = - \der_0^2 - \alpha^2 \left ( \der_1^2 - \der_2^2 \right )^2 - \beta^2 \der_1^2 \der_2^2 + \gamma^2 \der_3^2 \; ,  
\label{4.10} 
\ee
and $\widetilde \Pi_R (V)$ is shown in equation \eqref{3.4}. Note that $\Pi^2$ is symmetric under the exchange $\der_1 \leftrightarrow \der_2$.

\subsection{Computation rules}

The trace \eqref{4.9} is  computed by moving all the space-time derivatives on the right and the terms which do not contain derivatives  on the left so that 
\be
 \hbox{Tr}\,  \Bigl \{ F(x) \, G( i \der ) \, \Bigr \} = \int d^4 x \int {d^4 p \over (2 \pi)^4} \;  \hbox{Tr}\, \left \{  F(x) \, G(p) \right \} \; , 
 \label{4.11}
\ee
where  the correspondence $ i \der_\mu \rightarrow p_\mu$ has been used. Since $\Pi^2$ contains $p_0$ and $p_3$ at power 2, and $p_1$ and $p_2$ at power 4,  the integration over the momenta in the euclidean region  gives rise to the following powers of $\epsilon $
\be
\int d^4 p \, e^{ \epsilon \Pi^2 } [p_{0,3}]^a \, [p_{1,2}]^b  \sim \epsilon^{-3/2- a/2 - b/4} \; . 
\label{4.12} 
\ee
In the $\epsilon \rightarrow 0 $ limit,  expression \eqref{4.9} is a sum of a large number of nonvanishing contributions. Many of these contributions do not play a part  in the anomaly because they are just equal to the variation of local counterterms.  So, let us concentrate on the relevant (as far as the anomaly is concerned) terms which are of the type 
\be
\hbox{relevant terms} \sim \der_\mu \theta_R(x) \, V_\nu(x) \der_\tau V_\lambda (x)   \quad , \quad (\hbox {with ~ } \mu \not=\nu \not= \tau \not= \lambda )
\label{4.13}
\ee
in which there is not a couple of the indices $\mu , \nu , \tau  , \lambda $ which assume the same value.  Let  $\Gamma_{ct} [V] $ be the sum of the local counterterms whose gauge variation cancels precisely the integrable contributions of  $\delta_{\theta_R} \Gamma^{reg}_R [V]$ which are not of the type \eqref{4.13}. With the definition $\Gamma_R [V] = \Gamma_R^{reg}[V] - \Gamma_{ct} [V] $, we shall now consider the gauge variation of $\Gamma_R [V]$ in the $\epsilon \rightarrow 0 $ limit. 

There are $4! = 24 $ contributions of type \eqref{4.13}, which are contained in the $\epsilon^2$ term of the expansion \eqref{4.9}. 
In order to obtain a nonvanishing result in the $\epsilon \rightarrow 0 $ limit, one needs to compensate the powers of the cut-off $\epsilon$ by powers of the momenta in the integrals \eqref{4.11} and \eqref{4.12}. We will need to extract powers of the momenta also from the  
 exponential factors of the type $e^{ q\epsilon \Pi^2}$. More precisely, when one exponential factor $e^{ q\epsilon \Pi^2}$ commutes with a function $f(x)$, it gives the expression   
  \bea
[  e^{q \epsilon \Pi^2}, f(x) ] &=& - \epsilon q \Bigl \{  2 \der_0 f(x) \der_0 - 2 \gamma^2 \der_3 f(x) \der_3 + 4 \alpha^2 (\der_1 f(x) \der^3_1 + \der_2 f (x) \der^3_2)\nonumber \\ 
&& + 2( \beta^2 - 2\alpha^2) (\der_1  f (x) \der_1 \der^2_2 +  \der_2 f (x) \der^2_1 \der_2) \Bigr \} e^{q \epsilon \Pi^2} + \cdots 
\label{4.14}
  \eea
  where, in agreement with relation \eqref{4.12},  the first two terms give  rise to contributions of order $\epsilon^{1/2}$, 
  \be
-  \epsilon q \Bigl \{  2 \der_0 f(x) \der_0 - 2 \gamma^2 \der_3 f(x) \der_3 \Bigr \} \longrightarrow \epsilon^{1/2} \; , 
  \label{4.15}
  \ee
whereas   the remaining  two terms give rise to  contributions of order $\epsilon^{1/4}$
\be
-  \epsilon q \Bigl \{  4 \alpha^2 (\der_1 f(x) \der^3_1 + \der_2 f (x) \der^3_2) + 2( \beta^2 - 2 \alpha^2) (\der_1  f (x) \der_1 \der^2_2 +  \der_2 f (x) \der^2_1 \der_2) \Bigr \} \longrightarrow \epsilon^{1/4} \; , 
  \label{4.16}
\ee   
and the dots stand for terms which turn out to be irrelevant  (they produce vanishing outcomes  in the $\epsilon \rightarrow 0$ limit). 

\subsection{Addition of the contributions}

We have found 144 nonvanishing  contributions to $\delta_{\theta_R} \Gamma_R [V]$ and their sum can be written in the form 
\be
\delta_{\theta_R} \Gamma_R [V]= \frac{-i 8 \alpha \beta \gamma \epsilon^3}{3} \sum_n  c_n \int d^4 x \, F_n (x) \int \frac{d^4p}{(2 \pi)^4} e^{\epsilon \Pi^2 } \, G_n (p) \; , 
\label{4.17}
\ee
where the sum contains 24 addenda and the values of $c_n$, $F_n(x)$ and $G_n(p)$ are shown in Table \ref{T1}. 
   
\medskip

\begin{table}
\begin{tabular}{|c|c|c|c|c|c|c|}
 \hline
{\hskip  0.2 cm}  $c_n$ {\hskip 0.2 cm}  & {\hskip 1 cm}  $F_n(x)$ {\hskip 1 cm} & {\hskip 1 cm} $G_n(p)$  {\hskip 1 cm}  &  {\hskip 1 cm} & $c_n$ {\hskip 0.2 cm}  & {\hskip 1 cm}  $F_n(x)$ {\hskip 1 cm} & {\hskip 1 cm} $G_n(p)$  {\hskip 1 cm}\\ 
 \hline 
 \hline 
 $  1 $ & $ V_0 \der_1 \theta_R \der_2 V_3$ & $ \alpha^2 p_1^6 + (\beta^2 - \alpha^2) p_1^4 p_2^2  $ & \hspace{2cm} & $ 1 $ & $V_2 \der_0 \theta_R \der_1 V_3$ & $\alpha^2 p_1^6 + (\beta^2 - \alpha^2) p_1^4  p^2 $\\
\hline 
$ -2   \gamma^2 $ & $V_0 \der_1 \theta_R \der_3 V_2$ & $p_1^2 p_3^2  $ & & $ -2 \gamma^2 $ & $ V_2 \der_0 \theta_R \der_3 V_1$ & $ p_1^2 p_3^2  $ \\
\hline 
$ 2   \gamma^2 $ & $V_0 \der_2 \theta_R \der_3 V_1$ & $ p_1^2 p_3^2  $ & & $ 2 \gamma^2 $ & $V_2 \der_1 \theta_R \der_3 V_0$ & $ p_1^2 p_3^2  $ \\
\hline 
$   -1 $ & $V_0 \der_2 \theta_R \der_1 V_3$ & $ \alpha^2 p_1^6 + (\beta^2 - \alpha^2) p_1^4  p_2^2 $ & & $ 2 $ & $ V_2 \der_1 \theta_R \der_0 V_3$ & $ p_1^2 p_0^2  $ \\
\hline
$ 1 $ & $V_0 \der_3 \theta_R \der_1 V_2$ & $ \alpha^2 p_1^6 + (\beta^2 - \alpha^2) p_1^4  p_2^2 $ & & $ -2 $ & $V_2 \der_3 \theta_R \der_0 V_1$ & $p_1^2 p_0^2  $  \\
\hline  
$ -1 $ & $V_0 \der_3 \theta_R \der_2 V_1$ & $ \alpha^2 p_1^6 + (\beta^2 - \alpha^2) p_1^4  p_2^2 $ & & $ -1 $ & $ V_2 \der_3 \theta_R \der_1 V_0$ & $\alpha^2 p_1^6 + (\beta^2 - \alpha^2) p_1^4  p_2^2 $\\
\hline 
 $ -2 $ & $ V_1 \der_2 \theta_R \der_0 V_3$ & $ p_1^2 p_0^2  $ & & $ 1 $ & $ V_3 \der_2 \theta_R \der_1 V_0$ & $ \alpha^2 p_1^6 + (\beta^2 - \alpha^2) p_1^4  p_2^2 $  \\
\hline
$  -2 \gamma^2 $ & $ V_1 \der_2 \theta_R \der_3 V_0$ & $ p_1^2 p_3^2  $ & & $ 2 $ & $V_3 \der_2 \theta_R \der_0 V_1$ & $ p_1^2 p_0^2  $\\
\hline 
 $ 2 \gamma^2 $ & $V_1 \der_0 \theta_R \der_3 V_2$ & $ p_1^2 p_3^2  $ & &$ -2 $ & $V_3 \der_1 \theta_R \der_0 V_2$ & $ p_1^2 p_0^2  $  \\
\hline 
$ -1 $ & $V_1 \der_0 \theta_R \der_2 V_3$ &  $\alpha^2 p_1^6 + (\beta^2 - \alpha^2) p_1^4  p_2^2 $  & &  $ -1 $ & $V_3 \der_1 \theta_R \der_2 V_0$ & $ \alpha^2 p_1^6 + (\beta^2 - \alpha^2) p_1^4  p_2^2 $\\
\hline 
$ 1 $ & $ V_1 \der_3 \theta_R \der_2 V_0$ & $\alpha^2 p_1^6 + (\beta^2 - \alpha^2) p_1^4  p_2^2 $ & & $ 1 $ & $ V_3 \der_0 \theta_R \der_2 V_1$ & $\alpha^2 p_1^6 + (\beta^2 - \alpha^2) p_1^4  p_2^2 $ \\
\hline 
$2 $ & $V_1 \der_3 \theta_R \der_0 V_2$ & $ p_1^2 p_0^2  $ & & $ -1 $ & $ V_3 \der_0 \theta_R \der_1 V_2$ & $\alpha^2 p_1^6 + (\beta^2 - \alpha^2) p_1^4  p_2^2 $ \\
\hline 
\end{tabular}
\caption{Addenda in the sum of Eq. \eqref{4.17} .}
\label{T1}
\end{table}

\bigskip 

Let us recall that in momentum space ($ i \der_\mu \rightarrow p_\mu$) one has 
\be
e^{ \epsilon \Pi^2} = e^{ \epsilon p_0^2} \, e^{- \epsilon \gamma^2 p_3^2} \, e^{- \epsilon [\alpha^2 (p_1^2 - p_2^2)^2 + \beta^2 p_1^2 p_2^2]} \; . 
\label{4.18}
\ee
In agreement with the Feynman $\varepsilon$-convention of the  propagator,  the analytic continuation in the euclidean region is obtained according to   $p_0 \rightarrow i p_4$ with real $p_4$.  One gets 
\be
\int dp_4   \, e^{- \epsilon  p_4^2} = \pi^{1/2}  \, \epsilon^{-1/2} \quad , \quad \int dp_4   \, e^{- \epsilon  p_4^2} \, p_4^2 = \mezzo \, \pi^{1/2}\, \epsilon^{-3/2}  \; , 
\label{4.19}
\ee
and
\be
\int dp_3   \, e^{- \epsilon \gamma^2 p_3^2} = \pi^{1/2}  \, \epsilon^{-1/2}\, \frac{1}{ | \gamma | } \quad , \quad \int dp_3   \, e^{- \epsilon \gamma^2 p_3^2} \, p_3^2 = \mezzo \, \pi^{1/2}\, \epsilon^{-3/2} \, \frac {1}{ | \gamma |^3}  \; .  
\label{4.20}
\ee
Let us define 
\be
\int dp_1 \, dp_2 \; e^{- \epsilon [\alpha^2 (p_1^2 - p_2^2)^2 + \beta^2 p_1^2 p_2^2]} \; p_1^2 =  \pi \, \epsilon^{-1} \, X(\alpha ,\beta)\; .  
\label{4.21}
\ee
Since 
\bea
- {\der\over \der \epsilon }  \int dp_1 \, dp_2 \; e^{- \epsilon [\alpha^2 (p_1^2 - p_2^2)^2 + \beta^2 p_1^2 p_2^2]} \; p_1^2 &=& \nonumber \\
&& {\hskip -5 truecm} =  \int dp_1 \, dp_2 \; e^{- \epsilon [\alpha^2 (p_1^2 - p_2^2)^2 +\beta^2 p_1^2 p_2^2]} \left [ \alpha^2 p_1^6 + (\beta^2 - \alpha^2) p_1^4 p_2^2 \right ] \, , 
\label{4.22}
\eea
from equations (\ref{4.21}) and (\ref{4.22}) one derives 
\be
 \int dp_1 \, dp_2 \; e^{- \epsilon [\alpha^2 (p_1^2 - p_2^2)^2 +\beta^2 p_1^2 p_2^2]} \left [ \alpha^2 p_1^6 + (\beta^2 - \alpha^2) p_1^4 p_2^2 \right ] 
  =  \pi \, \epsilon^{-2} \, X(\alpha , \beta)\; .  
  \label{4.23}
 \ee
Therefore, the momenta integrals with appear in equation (\ref{4.17}) take the values 
\be
\int {d^4 p \over (2 \pi^4)} e^{ \epsilon \Pi^2 }\, p_1^2 p_3^2 \longrightarrow  \epsilon^{-3} \left ( {i \over 32 \pi^2}\right ) \frac{1}{|\gamma |^3 } X(\alpha , \beta )\; , 
\label{4.24}
\ee
\be
\int {d^4 p \over (2 \pi^4)} e^{ \epsilon \Pi^2 }\, p_1^2 p_0^2  \longrightarrow \epsilon^{-3} \left ( {- i \over 32 \pi^2}\right ) \frac{1}{|\gamma |} X(\alpha , \beta )\; , 
\label{4.25}
\ee
\be
\int {d^4 p \over (2 \pi^4)} e^{ \epsilon \Pi^2 }\, 
(\alpha^2 p_1^6 + (\beta^2 - \alpha^2) p_1^4  p_2^2) \longrightarrow  \epsilon^{-3} \left ( { i \over 16 \pi^2}\right ) \frac{1}{| \gamma |} X(\alpha , \beta )\; . 
\label{4.26}
\ee
Consequently, the sum \eqref{4.17} is given by 
\be
\delta_{\theta_R} \Gamma_R [V] =  - \frac{  \alpha \beta \gamma X(\alpha , \beta )}{6 \pi^2 | \gamma | } \int d^4 x  \, \epsilon^{\mu \nu \tau \lambda}   \der_\mu \theta_R (x)  V_\nu (x) \der_\tau V_\lambda (x) \; . 
\label{4.27}
\ee
Even if the fermion operator $\Pi_R(V)$ is not Lorentz covariant and differs  from the standard Dirac operator, and even if  $\Pi_R(V)$ contains dimensioned parameters, still  the sum of the various contributions to $\delta_{\theta_R} \Gamma_R [V]$ ---quite remarkably--- reproduces the standard form 
$$ 
F \wedge F = \left ( \der_\mu V_\nu - \der_\nu V_\mu \right ) \left ( \der_\lambda V_\tau - \der_\tau V_\lambda \right )  \, dx^\mu \wedge dx^\nu \wedge dx^\lambda \wedge dx^\tau 
$$ 
of the chiral anomaly \eqref{3.10}.  As it has been mentioned in section~\ref{general}, this is  a consequence of the fact that, in the abelian case,  the nontrivial solution of the cohomological problem \eqref{3.9} is given precisely by $F \wedge F$.  

\subsection{The final result}

Let us now derive the value of $X(\alpha , \beta)$. By means of a rescaling of the integration variables $p_i \rightarrow p_i / \epsilon^{1/4}$, equation \eqref{4.21} can be written as 
\be
X(\alpha ,\beta) = \frac{1}{2 \pi } \int dp_1 \, dp_2 \; e^{-  [\alpha^2 (p_1^2 + p_2^2)^2 + (\beta^2 - 4 \alpha^2 ) p_1^2 p_2^2]} \; (p_1^2 + p_2^2)  \, \; .  
\label{4.28}
\ee
By introducing two dimensional spherical coordinates, $p_1 = p \cos \varphi $, $p_2 = p \sin \varphi $,  one finds 
\be
\alpha^2 (p_1^2 + p_2^2)^2 + (\beta^2 - 4 \alpha^2 ) p_1^2 p_2^2 = \frac{ p^4}{8} \left [ (4 \alpha^2 + \beta^2)  + (4 \alpha^2 - \beta^2) \cos (4 \varphi )      \right ] \; ,  
\label{4.29}
\ee
and then 
\bea
X(\alpha ,\beta) &=& \frac{1}{8 \pi } \int_0^{2 \pi } d \varphi \; \frac {8}{4 \alpha^2 + 4 \beta^2 +   ( 4 \alpha^2 - \beta^2 ) \cos (4 \varphi) } \nonumber \\
&=& \frac{4}{\pi \sqrt{(4 \alpha^2 + \beta^2)^2 - (4\alpha^2 - \beta^2)^2}} \arctan \left [ \frac{(1 - \frac{4\alpha^2 - \beta^2 }{4\alpha^2 + \beta^2}) \, \tan (\theta/2)}{ \sqrt {1 - \left ( \frac{4\alpha^2 - \beta^2}{4\alpha^2 + \beta^2} \right )^2 }} \right ]_0^{\pi} \nonumber \\
&=& \frac{1}{2 \, |\alpha \beta |} \; .  
\label{4.30}
\eea
Equation \eqref{4.27} then reads 
\be
\delta_{\theta_R} \Gamma_R [V] =  - \frac{ 2 \alpha \beta \gamma}{ | \alpha  \beta \gamma |} \frac{1}{24 \pi^2 } \int d^4 x  \, \epsilon^{\mu \nu \tau \lambda}   \der_\mu \theta_R (x)  V_\nu (x) \der_\tau V_\lambda (x) \; . 
\label{4.31}
\ee
This means that, in equation \eqref{3.10}, the multiplicative factor $\cal N$ is given by 
\be
{\cal N} =   2 \, \frac{ \alpha \beta \gamma}{ | \alpha  \beta \gamma |} \; . 
\label{4.32}
\ee 
Therefore, in the double-Weyl model specified by the lagrangian \eqref{2.5}, the value of the axial anomaly \eqref{3.22} coincides with expression \eqref{2.9}. This concludes the perturbative quantum field theory proof of equations (\ref{2.9}) and (\ref{2.10}). 

\section{Nielsen-Ninomiya procedure}
\label{NN-procedure}

When the vector gauge invariance \eqref{2.2} is preserved,  in the presence of appropriate electric and magnetic fields,  the axial anomaly can be interpreted as the rate of production of ``particles  chirality'' as a consequence of a vacuum rearrangement \cite{nielsen1983} for the fermions. Nielsen and Ninomiya showed that the axial anomaly can be estimated by considering a simple quantum mechanical description of the system. In the presence of a uniform magnetic field along the $x^3$ direction, the spectrum of the hamiltonian associated with a Weyl cone displays a three-dimensional Landau level structure.
Among the gapped Landau level, there is a special gapless family of states with a chiral dispersion along the $x^3$ direction. This family of states is responsible for chiral anomaly upon the application of an electric field along $x^3$. Based on this idea, in the present section we shall rederive the result \eqref{2.9}. The main step will be to define the gapless and chiral families of states appearing when the Weyl cones are subject to a suitable magnetic field.   

\subsection{Classical external fields}

Let us consider the case in which  classical  electric and magnetic fields $E$ and  $B$ are directed in the $x^3$ direction,  
\be
A_0(x) = 0\quad , \quad A_1 (x) = 0 \quad , \quad  A_2 (x) =  - B x^1 \quad , \quad A_3(x) =    E x^0 \; ,  
\label{5.1} 
\ee
with $E > 0$ and $B > 0$. 
The equation of motion for the field  $\psi_R(x)$ takes the form 
\be
 \left \{ i \der_0 - \alpha \left [ \der_1^2 - (\der_2 - i B x^1)^2 \right ] \sigma^1 - \frac{\beta}{2} \left \{  \der_1 , (\der_2 - i B  x^1 ) \right \} \sigma^2 + i \gamma (\der_3 + i E x^0) \sigma^3 \right \} \psi_R ( x) = 0 \; . 
\label{5.2}
\ee
This equation of motion is translationally invariant along $x^2$ and $x^3$, so that we can consider, in full generality, wavefunctions with specific eigenvalues $k_2$ and $k_3$ of the two components $p_2$ and $p_3$ of the momentum:      
\be
 \psi_R ( x) = e^{i k_2 x^2} \, e^{i k_3 x^3} \, \chi (x^0, x^1) = e^{i k_2 x^2} \, e^{i k_3 x^3} \,
 \begin{pmatrix} \chi_\Up (x^0, x^1) \\ \chi_\Dn (x^0, x^1) \end{pmatrix}
 \; .  
 \label{5.3}
\ee
It is useful to introduce the operators 
\be
\eta = \frac{1}{\sqrt {2 B}} \, \der_1 + \sqrt {\frac{B}{2}} \left ( x^1 - \frac{k_2}{B}  \right )  \quad , \quad \eta^\dagger = - \frac{1}{\sqrt {2 B}} \, \der_1 + \sqrt {\frac{B}{2}} \left ( x^1 - \frac{k_2}{B}  \right ) \; , 
\label{5.4}
\ee
which satisfy the standard  commutation relation of  annihilation and creation operators,  $[\eta , \eta^\dagger] = 1$.
The associated  wavefunction
\begin{equation}
h_0 (x^1) = (B/\pi)^{1/4} \exp \left[-\frac{B}{2}(x^1-k_2/B)^2\right]
\end{equation} 
corresponds to the ground state of the 1D harmonic oscillator in the $x^1$ direction centered in $k_2/B$ \cite{landauQM} such that $\eta h_0=0$. Equation \eqref{5.2} allows us to define the hamiltonian of the system for the right modes:  
\be
i \der_0 \begin{pmatrix} \chi_\Up \\ \chi_\Dn \end{pmatrix} = \begin{pmatrix} \gamma (k_3 +  E x^0) & B(\alpha - \beta/2) \eta^2 + B (\alpha + \beta /2) {\eta^\dagger}^2 \\   B(\alpha + \beta/2) \eta^2 + B (\alpha - \beta /2) {\eta^\dagger}^2 & - \gamma (k_3 +  E x^0)
\end{pmatrix} \begin{pmatrix} \chi_\Up \\ \chi_\Dn \end{pmatrix} \equiv H_R(\gamma) \begin{pmatrix} \chi_\Up \\ \chi_\Dn \end{pmatrix}\; . 
\label{5.5}
\ee
The equation of motion for  the field $\psi_L(x)$
can be obtained from equation \eqref{5.2} by means of the substitution $\gamma \to -\gamma$. In particular $H_L(\gamma)=H_R(-\gamma)$ in such a way that for every eigenfunction $\psi_R(x,\gamma)$ of the hamiltonian $H_R$, there exists a corresponding eigenfunction $\psi_L(x,\gamma)= \psi_R(x,-\gamma)$ of the hamiltonian $H_L$.

\subsection{Normalizable zero-energy modes}

In order to  search for the chiral gapless Landau level which characterize the spectrum of $H_R(k_3)$, we must determine the zero-energy eigenstates of Eq. \eqref{5.5}. In particular, following Nielsen and Ninomiya, we consider the static problem with $E=0$ and we assume $\alpha \neq \pm \beta /2$.
Since the $\gamma$-term in $H_R$ anticommutes with the $\alpha$ and $\beta$ contribution, to search for a zero-energy solution we must impose $k_3=0$. Therefore we must look for normalized solutions $\chi_0$ of the equation:
\be
\begin{pmatrix} 0 & B(\alpha - \beta/2) \eta^2 + B (\alpha + \beta /2) {\eta^\dagger}^2 \\   B(\alpha + \beta/2) \eta^2 + B (\alpha - \beta /2) {\eta^\dagger}^2 & 0  
\end{pmatrix}  \begin{pmatrix} \chi_{0,\Up} \\ \chi_{0,\Dn} \end{pmatrix}  =0 \,.
\label{5.8}
\ee
Note that this equation is valid for both right $\psi_R(x)$ and left $\psi_L(x)$ field components, as long as $k_3=E=0$.
This equation can be recast into the following relations for the components $\chi_{0,\Up}$ and $\chi_{0,\Dn}$:
\begin{align}
&\left [ (\alpha - \beta/2) \eta^2 +  (\alpha + \beta /2) {\eta^\dagger}^2 \right ]  \chi_{0,\Dn} (\alpha,\beta,x_1) =0 \,, \label{zero1} \\
&\left [ (\alpha + \beta/2) \eta^2 +  (\alpha - \beta /2) {\eta^\dagger}^2 \right ]  \chi_{0,\Up} (\alpha,\beta,x_1) =0\;  \label{zero2}.
\end{align} 
These equations show that the two components are independent on each other, however we will show in the following that, given an arbitrary choice of $\alpha$ and $\beta$, only one of them can be normalized at a time. From the previous equations we also deduce that,  if $\chi_{0,\Up}(\alpha,\beta,x_1)$ is a solution of Eq. \eqref{zero2}, $\chi_{0,\Dn}(\alpha,\beta,x_1)=\chi_{0,\Up}(\alpha,-\beta,x_1)$ will be a solution of Eq. \eqref{zero1}, therefore we can limit our research to Eq. \eqref{zero2} without loss of generality. 

  To solve  equation \eqref{zero2}, let us consider the basis provided by the wavefunctions of the harmonic oscillator $h_n(x_1) = \left(\eta^{\dag}\right)^n h_0(x_1)/\sqrt{n!}$: the operator in Eq. \eqref{zero2} allows for transitions between $h_n$ and $h_{n \pm 2}$ only and the operator $(\alpha - \beta/2) \eta^2$ annihilates both $h_0$ and $n_1$. The wavefunctions $h_2$ and $h_3$, instead, cannot be annihilated by the operator in Eq. \eqref{zero2} and they must be absent from $\chi_{0,\Up}$. We deduce that there are two possible solutions of Eq. \eqref{zero2} given by suitable linear combinations of the wavefunctions $h_{4n}$ and $h_{4n+1}$ respectively:
\bea
{\chi}_{0,\Up}^{(1)}(x_1) &=& \sum_{n=0}^\infty (-1)^n d_n \left ( \frac{\eta^\dagger }{2} \right )^{4n}  h_0 (x_1) \; , \nonumber \\
{\chi}_{0,\Up}^{(2)}(x_1) &=& \sum_{n=0}^\infty (-1)^n e_n \left ( \frac{\eta^\dagger }{2} \right )^{4n} \eta^\dagger \,  h_0 (x_1)  \; , 
\label{5.9}
\eea
where $d_n$ and $e_n$ are real coefficients.   Equation \eqref{zero2} for the wavefunctions $\chi_{0,\Up}^{(i)}$ is fulfilled when: 
\bea
\frac{d_{n+1}}{d_n} &=& \left ( \frac{\alpha - \beta/2}{\alpha + \beta/2}\right ) \frac{1}{(n+1)(n+ 3/4)} \; , \nonumber \\
\frac{e_{n+1}}{e_n} &=& \left ( \frac{\alpha - \beta/2}{\alpha + \beta/2}\right ) \frac{1}{(n+5/4)(n+ 3/2)} \; .  
\label{5.10}
\eea
Both functions $\chi_{0,\Up}^{(i)}$ are normalizable if $\alpha\beta>0$. Indeed, one has for instance:  
\be
\int dx_1 \, |{\chi}_{0,\Up}^{(1)}(x_1)|^2 = \sum_n |d_n|^2 \left (\frac{1}{4} \right )^{4n} (4 n) ! \; .   
\label{5.12}
\ee 
The sum \eqref{5.12} is convergent since the large $n$ behaviour of the ratio of two consecutive addenda is given  by 
 \be
 \lim_{n \rightarrow \infty} \left [ |d_{n+1}|^2 \left (\frac{1}{4} \right )^{4(n+1)} [4 (n+1)] ! \right ]  
 \left [ |d_n|^2 \left (\frac{1}{4} \right )^{4n} (4 n) ! \right ]^{-1} 
 = \left ( \frac{\alpha - \beta/2}{\alpha + \beta/2}\right )^2 \; , 
 \label{5.13} 
 \ee 
and when $\alpha \beta >0 $ one has $ [ (\alpha - \beta/2) / (\alpha + \beta/2) ]^2 < 1$. Thus ${\chi}_{0,\Up}^{(1)}$ has finite norm. 
Similarly, one easily shows that ${\chi}_{0,\Up}^{(2)}$ is normalizable as well.  Moreover, the functions 
${\chi}_{0,\Up}^{(1)}(x_1)$ and ${\chi}_{0,\Up}^{(2)}(x_1)$ are orthogonal, because they are obtained by applying even or odd numbers of creation operators $\eta^\dagger$ respectively on the ground state wavefunction $h_0 (x_1)$. The previous functions, however, are not normalizable for $\alpha\beta <0$; this implies that, when the wavefunctions ${\chi}_{0,\Up}^{(i)}(\alpha,\beta)$ are well-defined, the corresponding wavefunctions $\chi_{0,\Dn}^{(i)}(\alpha,\beta) = \chi_{0,\Up}^{(i)}(\alpha,-\beta)$ are not. We conclude that, when $\alpha \beta >0 $ there are only two independent zero-energy modes given by:
 \be
\chi_0^{(1)} = \begin{pmatrix}   {\chi}_{0,\Up}^{(1)}(x_1) \\ 0 \end{pmatrix} \,, \qquad \chi_0^{(2)} = \begin{pmatrix}   {\chi}_{0,\Up}^{(2)}(x_1) \\ 0  \end{pmatrix} \, ,  \qquad (\hbox{ if~} \alpha \beta >0\, )  \; , 
\label{5.14}
\ee
which satisfy equation \eqref{5.8}.  When $\alpha \beta <0 $, instead, the role of $\chi_{0,\Up}$ and $\chi_{0,\Dn}$ is exchanged. Two normalizable functions $\chi_{0,\Dn}$ can be obtained from Eq. \eqref{zero2}. In this case the component $\chi_{0,\Dn}$ can be described by two series of the kind \eqref{5.9} by imposing:
\bea
\frac{d_{n+1}}{d_n} &=& \left ( \frac{\alpha + \beta/2}{\alpha - \beta/2}\right ) \frac{1}{(n+1)(n+ 3/4)} \; , \nonumber \\
\frac{e_{n+1}}{e_n} &=& \left ( \frac{\alpha + \beta/2}{\alpha - \beta/2}\right ) \frac{1}{(n+5/4)(n+ 3/2)} \; ;    
\label{5.15}
\eea
for $\alpha\beta<0$, the convergence of these series is verified because $[ (\alpha + \beta/2) / (\alpha - \beta/2) ]^2 < 1$ and the zero-energy modes result
 \be
\chi_0^{(1)} = \begin{pmatrix}  0 \\  {\chi}_{0,\Dn}^{(1)}(x_1)  \end{pmatrix} \, , \quad \chi_0^{(2)} = \begin{pmatrix}    0 \\ {\chi}_{0,\Dn}^{(2)}(x_1)  \end{pmatrix} \,,  \qquad (\hbox{ if~} \alpha \beta <0 )  \; , 
\label{5.17}
\ee
with $\chi_{0,\Dn}^{(i)}(\alpha,-\beta) = \chi_{0,\Up}^{(i)}(\alpha,\beta)$, relating the components in Eqs. \eqref{5.14} and \eqref{5.17}. 

 The result in \cite{li2016} for the particular case with rotational invariance, for $\alpha  = \pm \beta/2$,  can be  recovered  observing that in these cases only the first term in each series for $\chi_0^{(i)}$ is different from zero. 

\subsection{Dirac sea and the axial anomaly}

So far we discussed the zero-energy case. Now we reintroduce the $\gamma$ term in \eqref{5.5} and we assume $k_3,E \neq 0$.
Let us consider the case $\alpha\beta >0$. The resulting chiral modes are of the form:
\begin{equation}
\psi^{(i)}_{R/L} (x) =  e^{i k_2 x^2} \, e^{i k_3 x^3} \, \chi_0^{(i)} (x^1)  f_{R/L} (x^0) \; , \label{5.18}
\end{equation}
with $\chi_0^{(i)} (x^1)$ defined in equation  \eqref{5.14} and the phase $f_{R/L}(x^0)$ to be determined by solving the Schroedinger equation  \eqref{5.5}: $i\partial_0 \psi_{R/L}(x^0) = H_{R/L}(x^0) \psi_{R/L}(x^0)$. This equation can be solved by considering that $\chi_0^{(i)}$ has only one component which is annihilated by the off-diagonal terms of $H_{R/L}$, whereas the time-dependent diagonal term implies:
\begin{equation}
f_R (x^0) = \exp\left[-i \gamma \left(k_3x^0 + \frac{E\left(x^0\right)^2}{2}\right)\right]\,, \qquad f_L (x^0) = \exp\left[i \gamma \left(k_3x^0 + \frac{E\left(x^0\right)^2}{2}\right)\right] ,  \qquad (\hbox{ if~} \alpha \beta >0 ) \,.
\label{f1}
\end{equation}
In the case $\alpha\beta<0$, instead, the wavefunctions $\chi_0^{(i)}$ possess only the second component (see equation \eqref{5.17}) and the resulting time dependence is:
\begin{equation}
f_R (x^0) = \exp\left[i \gamma \left(k_3x^0 + \frac{E\left(x^0\right)^2}{2}\right)\right]\,, \qquad f_L (x^0) = \exp\left[-i \gamma \left(k_3x^0 + \frac{E\left(x^0\right)^2}{2}\right)\right] ,  \qquad (\hbox{ if~} \alpha \beta <0 ) \,.
\label{f2}
\end{equation}   
With the definition of $\omega $ given by $i \der_0 \psi (x) = \omega \psi (x) $, we derive: 
\be
\omega_R = \gamma  (k_3 +  E x^0) \,, \qquad \omega_L =    - \gamma   (k_3 + E x^0) 
\, ,  \qquad (\hbox{ if~} \alpha \beta >0 ) \,,
\label{5.19}
\ee
and
\be
\omega_R = - \gamma  (k_3 +  E x^0) \,, \qquad \omega_L =  \gamma  (k_3 + E x^0)
\, ,  \qquad (\hbox{ if~} \alpha \beta <0 )  \, . 
\label{5.20}
\ee
These equations are consistent with a constant acceleration of the particles along $x^3$ given by $\partial_0 \omega$ and define the chiral nature of these gapless modes, reflecting the linear dispersion as a function of $k_3$. 
Vacuum stability requires that the values of $\omega_R $ and $\omega_L$ must be nonnegative.  Therefore, in agreement with the Dirac sea interpretation of the fermions ground state,     the stable vacuum  of the  system  corresponds to the state in which all the single-particle states with negative frequencies are occupied.  All the other Landau levels $\psi_R(x)$ and $\psi_L(x)$, which are orthogonal to the modes \eqref{5.18}, are not chiral, namely they have frequencies with a symmetric dispersion for $k_3 \to -k_3$ and they never cross zero energy: during the time evolution, the sign of their frequency does not change. This implies that these gapped Landau levels are either totally empty or totally filled, and, when considering the effect of the acceleration caused by $E$, they do not contribute to the net rate of change of the right or left particle number \cite{nielsen1983}. 
Thus, as far as the axial anomaly is concerned, we only need to discuss  the vacuum stability with respect to the modes \eqref{5.18}. 

Let us consider the case  $ \alpha \beta >  0$ and $\gamma >0$. The value \eqref{5.19} of $\omega_R $ is negative for $ k_3 <   - E x^0$. Therefore all the right-handed one-particle states with $k_3 <  -E x^0 $ must be occupied, and the available states for a $\psi_R$ particle are only those with $k_3 > -E x^0$.  Similarly,  since $\omega_L$ is negative for $k_3 > - E x^0$, all the left-handed one-particle states with  $k_3 > -  E x^0$ must be occupied and the available states for a $\psi_L$ particle are only those with $k_3 < -E x^0$. 

Let us recall that, if the one-particle states are labelled by the values $\ner k$ of the   momentum, the number $N$ of available states for one particle moving  inside a cubic box of volume $V= L^3$ is specified (in the large $V$ limit) by  $dN = L^3 d^3 k / (2 \pi)^3$.  Therefore, in our system the number $N_R$ of available $\psi_R$ states is determined by the product 
\be
N_R = 2 \times \hbox{Landau degeneracy} \times \hbox{range of } k_3 \; , 
\label{5.21}
\ee
where the factor $2$ is due to the presence of the two modes  $\chi_0^{(1)}$ and $ \chi_0^{(2)}$.  
The Landau degeneracy is determined by the range of $k_2$ which guarantees the particle localisation inside the box, $0 \leq ( x^1 - k_2 / B)  \leq L$,  
\be
\hbox{Landau degeneracy} = L \int_{0}^{BL} \frac{dk_2}{(2 \pi)} = \frac{BL^2}{2 \pi}  \; .
\label{5.22} 
\ee
One gets  
\be
N_R = 2 \times L \int_{0}^{BL} \frac{dk_2}{(2 \pi)} \times L \int_{-Ex^0}^{\infty} \frac{dk_3}{(2 \pi)}  = \frac{2 B V}{(2 \pi)^2} \int_{-Ex^0}^{\infty} dk_3 \; ,  
\label{5.23}
\ee
and then 
\be
\der_0 N_R = \frac{2 EB V}{4 \pi^2} \; . 
\label{5.24}
\ee
By means of the same argument, one determines the number $N_L$ of available $\psi_L$ states  
\be
N_L = 2 \times L \int_{0}^{BL} \frac{dk_2}{(2 \pi)} \times L \int_{-\infty}^{-Ex^0} \frac{dk_3}{(2 \pi)}  = \frac{2 B V}{(2 \pi)^2} \int_{-\infty }^{- Ex^0} dk_3 \; ,   
\label{5.25}
\ee
and thus 
\be
\der_0 N_L = - \frac{2 EB V}{4 \pi^2} \; . 
\label{5.26}
\ee
Consequently, one finds 
\be
  \int d^3 x \, \der_0 J_A^0 (x) = \der_0 N_R - \der_0 N_L =  \frac{ V 4 EB}{ 4 \pi^2} =   -   V \frac {2}{4 \pi^2} \epsilon^{\mu \nu \tau \lambda} \der_\mu A_\nu \der_\tau A_\lambda  \, , 
\label{5.27}
\ee 
which is in agreement with the perturbative computation of the axial anomaly of section~\ref{perturb}. It is now easy to verify that, for all the possible nontrivial values of $\alpha, \beta$ and $\gamma$, the axial anomaly computed by means of the Nielsen-Ninomiya method  coincides with expression \eqref{2.9}. This concludes the derivation of the result  \eqref{2.9} by means of the Nielsen-Ninomiya method. 

Finally we observe that the vector current is conserved, since 
\be
 \int d^3 x \left ( \der_0 J^0_R (x) + \der_0 J^0_L(x) \right ) = \der_0 N_R + \der_0 N_L =  0   \; .  
\label{5.28}
\ee

{\color{black} 
The Nielsen-Ninomiya method suggests that the axial anomaly is stable against perturbations of the chemical potentials around the (zero) energy of the multi-Weyl nodes. 
We shall discuss this issue in Section~\ref{Stability} }
   
\section{Atiyah-Singer index}
\label{index}

The axial anomaly can also be interpreted \cite{AS,BERT,CALL,ALVA,FRIE} as the index of the  euclidean analytic extension of the operator which acts on the fermion field in the expression \eqref{2.5} of the lagrangian.  The index can be defined as the number of nontrivial normalizable solutions with zero eigenvalues of this lagrangian operator, having support in ${\mathbb R}^4$. Moreover, the null solutions for the right and left parts can be identified separately, and the index results from the difference of their numbers. By using the Atiyah-Singer approach, in this section we shall rederive the result \eqref{2.9}.  

\subsection{Particles and antiparticles}

In the presence of the classical  external electric $E$ and magnetic $B$ fields shown in equation \eqref{5.1},  from the lagrangian \eqref{2.5}  one can derive    the equations of motion for the ``right-handed'' particle wave functions $\psi_R $,  
\be
\left \{     i \der_0 - \alpha \left [ \der_1^2 - (\der_2 - i B x^1)^2 \right ] \sigma^1 - \frac{\beta}{2} \left \{  \der_1 , (\der_2 - i B  x^1 ) \right \} \sigma^2 + i \gamma (\der_3 + i E x^0) \sigma^3 \right \} \psi_R(x) = 0 \; , 
\label{6.1}
\ee
and for the ``left-handed'' particle wave functions $\psi_L$, 
\be
\left \{  i \der_0 - \alpha \left [ \der_1^2 - (\der_2 - i B x^1)^2 \right ] \sigma^1 - \frac{\beta}{2} \left \{  \der_1 , (\der_2 - i B  x^1 ) \right \} \sigma^2 - i \gamma (\der_3 + i E x^0) \sigma^3  \right \} \psi_L(x) = 0 \; . 
\label{6.2}
\ee
Let $\psi_R^C$  and $\psi_L^C$ represent the wave functions  of the ``right-handed''  and ``left-handed''  antiparticles respectively,  
\be
\psi_R^C = \left ( \psi_R \right )^C = (-i \sigma^2) \, \psi_R^*(x) \qquad , \qquad  \psi_L^C = \left ( \psi_L \right )^C  = (i \sigma^2) \, \psi_L^*(x) \; . 
\label{6.3}
\ee 
The equations of motion for $\psi_R^C$  and $\psi_L^C$ in the given classical electromagnetic background \eqref{5.1} take the form 
\be
\left \{   -  i \der_0 + \alpha \left [ \der_1^2 - (\der_2 + i B x^1)^2 \right ] \sigma^1 + \frac{\beta}{2} \left \{  \der_1 , (\der_2 + i B  x^1 ) \right \} \sigma^2 + i \gamma (\der_3 - i E x^0) \sigma^3 \right \} \psi_R^C(x) = 0  \; , 
\label{6.4}
\ee
and
\be
\left \{  -i \der_0 + \alpha \left [ \der_1^2 - (\der_2 + i B x^1)^2 \right ] \sigma^1 + \frac{\beta}{2} \left \{  \der_1 , (\der_2 + i B  x^1 ) \right \} \sigma^2 - i \gamma (\der_3 - i E x^0) \sigma^3  \right \} \psi_L^C (x) = 0 \; . 
\label{6.5}
\ee
The  field operators $\psi_R$ and $\psi_L$ describe four kinds of particles: one ``right-handed'' particle and its antiparticle, and one ``left-handed'' particle and its antiparticle. The four relations \eqref{6.1},  \eqref{6.2}, \eqref{6.4} and \eqref{6.5} represent precisely a complete set of corresponding equations. 

\subsection{Euclidean zero modes}

With fixed electromagnetic background, let us consider the analytic extension of the operators entering the equations of motion \eqref{5.1}, \eqref{5.2}, \eqref{5.4} and \eqref{5.5}  in the euclidean region \cite{SY} (which is obtained by means of the replacement $p_0 \rightarrow i p_0$).   We need to determine \cite{AS} the corresponding normalizable zero modes in ${\mathbb R}^4$.   

Let us examine the case $\alpha \beta >0$ and $\gamma>0$.   One can specify the values  of the two spatial  components $p_2$ and $p_3$ of the momentum by putting
\bea
\psi_R(x) &=& e^{ik_2x^2} e^{i k_3 x^3} \tilde{\psi}_R(x^1 , x^0) \,, \qquad \psi_L(x) = e^{ik_2x^2} e^{i k_3 x^3} \tilde{\psi}_L(x^1 , x^0) \nonumber \\
\psi^C_R(x) &=& e^{-ik_2x^2} e^{-i k_3 x^3} \tilde{\psi}^C_R(x^1 , x^0) \, , \qquad \psi^C_L(x) = e^{-ik_2x^2} e^{-i k_3 x^3} \tilde{\psi}^C_L(x^1 , x^0) \; . 
\label{6.6}
\eea 
In addition to the $\eta$ and $\eta^\dagger $ operators defined in equation \eqref{5.4}, it is useful to introduce the ladder operators $\zeta $ and $\zeta^\dagger $, 
\be
\zeta = \frac{1}{\sqrt {2 \gamma E}} \, \der_0 + \sqrt {\frac{\gamma E}{2}} \left ( x^0 + \frac{k_3}{E}  \right )  \quad , \quad \zeta^\dagger = - \frac{1}{\sqrt {2 \gamma E}} \, \der_0 + \sqrt {\frac{\gamma E}{2}} \left ( x^0 + \frac{k_3}{E}  \right ) \; ,  
\label{6.7}
\ee
satisfying the canonical commutation relations $[\zeta , \zeta^\dagger ] = 1$. Let $f_0 (x^0)$ be the normalised ground state wave function satisfying $\zeta f_0 (x^0) = 0 $. 
The euclidean analytic extensions of  equations \eqref{6.1}, \eqref{6.2}, \eqref{6.4} and \eqref{6.5} assume the form 
\be
\left \{    - \sqrt{\frac{\gamma E}{2}} (\zeta - \zeta^\dagger ) -  \alpha B (\eta^2 + (\eta^\dagger )^2 ) \sigma^1 + i \frac{\beta}{2}  B (\eta^2 - (\eta^\dagger )^2 )  \sigma^2   -  \sqrt{\frac{\gamma E}{2}} (\zeta + \zeta^\dagger )  \sigma^3 \right \} \widetilde \psi_R = 0 \; , 
\label{6.8}
\ee
\be
\left \{    - \sqrt{\frac{\gamma E}{2}} (\zeta - \zeta^\dagger ) -  \alpha B (\eta^2 + (\eta^\dagger )^2 ) \sigma^1 + i \frac{\beta}{2}  B (\eta^2 - (\eta^\dagger )^2 )  \sigma^2   +   \sqrt{\frac{\gamma E}{2}} (\zeta + \zeta^\dagger )  \sigma^3 \right \} \widetilde \psi_L = 0 \; , 
\label{6.9}
\ee
\be
\left \{     \sqrt{\frac{\gamma E}{2}} (\zeta - \zeta^\dagger ) +  \alpha B (\eta^2 + (\eta^\dagger )^2 ) \sigma^1 + i \frac{\beta}{2}  B (\eta^2 - (\eta^\dagger )^2 )  \sigma^2   +  \sqrt{\frac{\gamma E}{2}} (\zeta + \zeta^\dagger )  \sigma^3 \right \}  \widetilde \psi^C_R = 0 \; , 
\label{6.10}
\ee
\be
\left \{     \sqrt{\frac{\gamma E}{2}} (\zeta - \zeta^\dagger ) +  \alpha B (\eta^2 + (\eta^\dagger )^2 ) \sigma^1 + i \frac{\beta}{2}  B (\eta^2 - (\eta^\dagger )^2 )  \sigma^2   -  \sqrt{\frac{\gamma E}{2}} (\zeta + \zeta^\dagger )  \sigma^3 \right \} \widetilde \psi^C_L = 0 \; .  
\label{6.11}
\ee
The complete set of normalizable solutions in ${\mathbb R}^4$ of equations (\ref{6.8})-(\ref{6.11}) can easily be determined because these equations depend on  separated variable,  since $\eta $ and $\eta^\dagger $ commute with $\zeta$ and $\zeta^\dagger$. 
Equations (\ref{6.9}) and (\ref{6.10}) do not admit normalizable solutions. Whereas equation \eqref{6.8} admits the normalizable solutions  
\be
\widetilde \psi_R (x^1 , x^0 ) =   \begin{pmatrix}   \chi_{0,\Up}^{(i)}(x^1) \, f_0 (x^0) \\ 0 \end{pmatrix} \; , \qquad \hbox{ with } i = 1,2 \; ,  
\label{6.12}
\ee
and  equation \eqref{6.11} admits the normalizable solutions 
\be
\widetilde \psi^C_L (x^1 , x^0 ) =   \begin{pmatrix}  0 \\  \chi_{0,\Up}^{(i)}(x^1) \, f_0 (x^0)  \end{pmatrix} \; , \qquad \hbox{ with } i = 1,2 \;  ,  
\label{6.13}
\ee
where the functions $\chi_{0,\Up}^{(i)}$ are defined in equations \eqref{5.9} and we are exploiting the mapping between right and left sectors. By exploiting the separability of the Landau level structures defined by the $\zeta$ and $\eta$ operators, we can map the previous wavefunction in a 4D quantum Hall problem \cite{price2015,lohse18}. We obtain that inside a hypercube in ${\mathbb R}^4$ of hypervolume $V_4 = L^4$ the Landau degeneracy of the each of the zero modes \eqref{6.12} and \eqref{6.13} is given by 
\be
\hbox{Landau degeneracy} = L \int_{0}^{BL} \frac{dk_2}{(2 \pi)} \times L \int_0^{LE } \frac{dk_3}{(2 \pi)}= \frac{EBL^4}{4 \pi^2}  \; .
\label{6.14} 
\ee
Therefore the number $\nu_R$ of euclidean zero modes, which are associated with  the ``right-handed particles'',   is given by 
\be
\nu_R = 2 \times \frac{EBL^4}{4 \pi^2} \; , 
\label{6.15}
\ee
and the number $\overline \nu_L$ of euclidean zero modes, which are associated with the ``left-handed antiparticles'', is found to be 
 \be
\overline \nu_L = 2 \times \frac{EBL^4}{4 \pi^2} \; , 
\label{6.16}
\ee
Therefore the index of the euclidean extension of the lagrangian operator ---acting on the fermion fields--- turns out to be 
\be
\frac{\nu_R + \overline \nu_L}{V_{4}} = 4 \times \frac{EB}{4 \pi^2} = - \frac{2}{4 \pi^2} \epsilon^{\mu \nu \tau \lambda } \der_\mu A_\nu \der_\tau A_\lambda  \; ,  
\label{6.17}
\ee
which is in agreement with the expression \eqref{2.9} of the axial anomaly. One can easily verify that this agreement still holds for arbitrary values of $\alpha, \beta$ and $\gamma$. This concludes the rederivation of the result \eqref{2.9} by means of the Atiyah-Singer index argument. 

 We note that the approach followed in the present Section  suggests a direct relation between the lagrangian zero modes and the chiral states derived from the corresponding hamiltonians in Section \ref{NN-procedure}, allowing for a parallelism between the two related methods to obtain the chiral anomalies.  
 
\section{ Axial anomaly for triple-Weyl semimetals} 
\label{3-Weyl}

In this section, the axial anomaly for the triple-Weyl semimetals model  is derived by means of the Nielsen-Ninomiya  and  the Atiyah-Singer arguments because,  in this case, the perturbative quantum field theory procedure requires considerable effort.    
 
As it has been discussed in section~\ref{NN-procedure}, in order to implement the Nielsen-Ninomiya procedure we need to consider the equations of motion which are derived from the lagrangian \eqref{2.11} in the presence of the gauge fields background \eqref{5.1}.  For the moment, let us consider the case in which    $\gamma > 0 $. The analogues of equations \eqref{5.5} take the form 
\be
i \der_0 \begin{pmatrix} \chi_\Up \\ \chi_\Dn \end{pmatrix} = \begin{pmatrix}   \gamma (k_3 +  E x^0) & B(\alpha - \beta) \eta^3 + B (\alpha + \beta ) {\eta^\dagger}^3 \\   B(\alpha + \beta) \eta^3 + B (\alpha - \beta) {\eta^\dagger}^3 & - \gamma (k_3 +  E x^0)
\end{pmatrix} \begin{pmatrix} \chi_\Up \\ \chi_\Dn \end{pmatrix} \; ,  
\label{7.1}
\ee
Therefore, we search again zero-energy ($\omega =0$) normalizable solutions of the equation for $\gamma=0$:
\be
\begin{pmatrix} 0 & B(\alpha - \beta) \eta^3 + B (\alpha + \beta) {\eta^\dagger}^3 \\   B(\alpha + \beta) \eta^3 + B (\alpha - \beta ) {\eta^\dagger}^3 & 0 \end{pmatrix}  
\begin{pmatrix} \chi_\Up \\ \chi_\Dn \end{pmatrix} = 0 \; .  
\label{7.3}
\ee
For this purpose, similarly to equation \eqref{5.9}, we introduce the following ansatz:
\beq
\begin{pmatrix} \chi_\Up \\ \chi_\Dn \end{pmatrix} = \begin{pmatrix} \sum_n a_n \, h_n(x^1) \\ \sum_n b_n \, h_n(x^1) \end{pmatrix} \, ,
\label{7.4}
\eeq
where $h_n(x^1)$ are the wavefunction of the $n$-th Landau level. By inserting this expansion in Eq. \eqref{7.3} we obtain, for $n \geq 3$:
\begin{align}
& (\alpha + \beta) \, b_{n-3} \, K_{n-3,+} + (\alpha - \beta) \,  b_{n+3} \, K_{n+3,-} = 0  \label{7.5a}\\ 
& (\alpha - \beta) \, a_{n-3} \, K_{n-3,+} + (\alpha + \beta) \,  a_{n+3} \, K_{n+3,-} = 0
\label{7.5}
\end{align} 
where $K_{n+3,-}  = \sqrt{(n+1) (n+2) (n+3)}$ and $K_{n- 3, +}  = \sqrt{n (n-1) (n -2)}$.  We impose  $b_{n-3} = K_{n-3,+} = 0$ if $n<3$ . 

The two equations \eqref{7.5a} and \eqref{7.5} decouple. Therefore, by following the same argument presented in section~\ref{NN-procedure} after equation \eqref{5.13}, we conclude that normalizable solutions with $\omega = 0$ exist only if $\{a_n\} = 0$ or $\{ b_n \} = 0$. By  direct inspection, we find that   $\{a_n\} = 0$ if $\alpha \beta<0 $, while $\{b_n\} = 0$ if $\alpha \beta >0 $. In each case, we obtain three independent normalizable solutions with $\omega = 0$, corresponding to values $\{a_0, a_1,a_2\}$ or $\{b_0, b_1,b_2\}$ to be fixed. 
Following the same arguments of the Section  \ref{NN-procedure} one obtains: 
\be
  \int d^3 x \, \der_0 J_A^0 (x) = \der_0 N_R - \der_0 N_L =  \frac{3 EB V}{4 \pi^2} - \left (  -        \frac{3 EB V}{4 \pi^2} \right ) = 
  \frac{ V 6 EB}{ 4 \pi^2} =   -   V \frac {3}{4 \pi^2} \epsilon^{\mu \nu \tau \lambda} \der_\mu A_\nu \der_\tau A_\lambda    \; , 
\label{7.6}
\ee 
which is in agreement with equation \eqref{2.12}. One can easily verify that a modification of the signs of the coefficients $\alpha $, $\beta $ and $\gamma $ is taken into account  by the $\Theta ( \alpha , \beta , \gamma )$ factor defined  in expression \eqref{2.10}. \\

The derivation of the axial anomaly by means of the Atiyah-Singer argument is quite simple. Indeed, the counting of the euclidean zero modes is carried out by means of  two steps:  

\begin{itemize}
\item (by adopting the notations of section~\ref{index}), the  $\zeta ,  \zeta^{\dagger}$   part in the ${\mathbb R}^4$ euclidean space gives 
a factor 1, precisely as the case of the double-Weyl model; 
\item  whereas the $\eta , \eta^{\dagger}$ part gives a multiplicative factor 3,  because there are three independent normalizable solutions of equation \eqref{7.3} with ${\cal E} =0$.  
\end{itemize}
Consequently,  in agreement with equation \eqref{7.6}, the axial anomaly of the triple-Weyl model  reads 
 \beq
 \partial_{\mu} J_A^{\mu} (x)=   -  3 \, \Theta (\alpha , \beta , \gamma ) \frac{1}{4 \pi^2} \, \epsilon^{\mu \nu \tau \lambda} \der_\mu  A_\nu (x) \der_\tau A_\lambda (x)  \, .  
 \label{7.7}
\eeq
 This concludes the proof of equation \eqref{2.12}. 

\section{Triple-point semimetals model}
\label{3-point}

The lagrangian of the triple-point semimetals   model is shown in equation \ref{2.13}.  
In order to compute the axial anomaly, we shall first use the Nielsen-Ninomiya method,  and then the Atiyah-Singer argument,  because the standard perturbative approach suffers from difficulties due to the existence of  singular points in the parameter space. Let us consider the fermionic fields in the presence of the gauge background 
 $$
 A_0(x) = 0\quad , \quad A_1 (x) = 0 \quad , \quad  A_2 (x) =  - B x^1 \quad , \quad A_3(x) =    E x^0 \; ,  
 $$
 with $E > 0$ and $B > 0$.  Since the gauge background does not depend on $x^2$ and $x^3$, one can specify the values $k_2$ and $k_3$ of the components $p_2$ and $p_3$ of the momentum 
 \be
 \psi_R ( x) = e^{i k_2 x^2} \, e^{i k_3 x^3} \, \chi (x^0, x^1)  \qquad , \qquad \psi_L ( x) = e^{i k_2 x^2} \, e^{i k_3 x^3} \, \xi (x^0, x^1) 
 \; ,   
 \label{8.1}
\ee
and the equations of motion following from the lagrangian \eqref{2.13} take the form 
\bea
i \der_0 \chi &=& iv \left [ M_1 \der_1 +  M_2 (i k_2 - iB x^1 ) +  M_3 ( i k_3 + i E x^0 ) \right ] \chi \nonumber  \\ 
i \der_0 \xi &=& iv \left [ M_1 \der_1 +  M_2 (i k_2 - iB x^1 ) -  M_3 ( i k_3 + i E x^0 ) \right ] \xi \; . 
\label{8.2}
\eea
Let us recall  that we need to determine \cite{nielsen1983} the crossing rate  of  the energy eigenvalues of the single particle states through the zero level.   Since the (linear) time dependence of the hamiltonian is contained in the covariant derivative $D_3$ exclusively, it is convenient to introduce the two normalised eigenvectors $\Sigma_{\pm}$ of $M_3$ with nontrivial eigenvalues, 
 \be
\Sigma_{\pm} = \frac{1}{\sqrt 2}  \begin{pmatrix}  \mp  \, e^{i\theta/2} \\  i e^{-i\theta/2} \\   0 
 \end{pmatrix}  \qquad , \qquad M_3 \Sigma_{\pm} = \pm \Sigma_{\pm}   \; . 
 \label{8.3}
 \ee  
A normalizable (in the $x^1$ variable) zero eigenvector $u_0 (x^1)$    of  the ``reduced  hamiltonian''  $i v \left ( M_1 D_1 + M_2 D_2 \right )$ must satisfy the equation   
\be
 \left (  M_1 D_1 + M_2 D_2 \right ) u_0 = \begin{pmatrix} 0 & 0 &  e^{-i \theta} (Bx^1 - k_2) \\ 0 & 0 & e^{i \theta} ( i \der_1) \\ 
 -e^{i \theta}  (B x^1-k_2) & e^{-i \theta} ( -i \der_1 ) & 0 \end{pmatrix} u_0 (x^1) =0  \; . 
 \label{8.4}
\ee
The normalizable solutions of equation \eqref{8.4} are given by  
\bea
u_0(x^1) &=& \Sigma_+\, \exp \left [ - \frac{1}{2B} \left ( Bx^1 - k_2 \right )^2 e^{i 3 \theta}   \right ] \; \quad , \quad \hbox{ when } \quad  \cos (3 \theta ) > 0 \; , \nonumber \\
u_0(x^1) &=& \Sigma_- \, \exp \left [  \frac{1}{2B} \left ( Bx^1 - k_2 \right )^2 e^{i 3 \theta}   \right ] \; \quad , \quad \hbox{ when } \quad \cos (3 \theta ) < 0 \; .  
\label{8.5}
\eea
In the case $\cos ( 3 \theta ) = 0 $, equation \eqref{8.4} does not admit normalizable solutions.  We point out  that
this condition corresponds to the condition $\sin ( 3 \theta ) = 0 $ in the notation of \cite{bradlyn2016}.
The fermionic modes 
\be
\psi_R (x) =  e^{i k_2 x^2} \, e^{i k_3 x^3} \, u_0 (x^1) f_R (x^0)\qquad  , \qquad 
\psi_L (x) =  e^{i k_2 x^2} \, e^{i k_3 x^3} \, u_0 (x^1) f_L (x^0) \; , 
\label{8.6} 
\ee
 satisfy the equation $i \der_0 \psi (x) = \omega \psi (x) $,  \big(then  $ f_{R/L} (x^0)$ defined similarly as in \eqref{f1} and \eqref{f2}\big), with frequencies 
\be
\omega_R = -v  (k_3 +  E x^0) \qquad , \qquad \omega_L =    v   (k_3 + E x^0) 
\qquad ,  \qquad (\hbox{ if~} \cos (3 \theta ) >  0\, ) 
\; , 
\label{8.7}
\ee
and 
\be
\omega_R = v  (k_3 +  E x^0) \qquad , \qquad \omega_L =    - v   (k_3 + E x^0) 
\qquad ,  \qquad (\hbox{ if~} \cos (3 \theta ) <  0\, )  \; .  
\label{8.8}
\ee
During the time evolution, the values of these frequencies crosses the zero value; thus the modes \eqref{8.6} contribute to the axial anomaly. While the remaining  fermionic modes have frequencies with fixed signs and can be neglected \cite{nielsen1983}. Therefore, when $\cos (3 \theta ) > 0 $,  for particles moving inside a cubic box of volume $V=L^3$,  the numbers $N_R$ and $N_L$  of available one particle states  are given by  
\be
N_R =  L \int_{0}^{BL} \frac{dk_2}{(2 \pi)} \times L  \int_{-\infty }^ {-Ex^0} \frac{dk_3}{(2 \pi)}   \quad , \quad N_L =    L \int_{0}^{BL} \frac{dk_2}{(2 \pi)} \times L \int_{-Ex^0}^{\infty} \frac{dk_3}{(2 \pi)} \; . 
\label{8.9}
\ee
Hence the vector gauge symmetry is not anomalous 
\be
\der_0 N_R + \der_0 N_L = 0 \; , 
\label{8.10}
\ee
whereas 
\be
  \int d^3 x \, \der_0 J_A^0 (x) = \der_0 N_R - \der_0 N_L =  - \frac{ V 2 EB}{ 4 \pi^2} =      V \frac {1}{4 \pi^2} \epsilon^{\mu \nu \tau \lambda} \der_\mu A_\nu \der_\tau A_\lambda    \; .  
\label{8.11}
\ee 
Similarly, when $\cos (3 \theta  ) < 0$, one finds 
\be
  \int d^3 x \, \der_0 J_A^0 (x) = \der_0 N_R - \der_0 N_L =  + \frac{ V 2 EB}{ 4 \pi^2} =     - V \frac {1}{4 \pi^2} \epsilon^{\mu \nu \tau \lambda} \der_\mu A_\nu \der_\tau A_\lambda    \; .  
\label{8.12}
\ee 

As illustrated in section~\ref{index}, the computation of the axial anomaly by means of the Atiyah-Singer approach  is strictly connected with the Nielsen-Ninomiya method. Indeed, in the presence of the gauge background \eqref{5.1}, the number of euclidean zero modes can be written as the product of the  
Landau degeneracy \eqref{6.14} with the number of the normalised zero modes of the ``reduced hamiltonian'' of equation \eqref{8.4}.   
Therefore, also for the triple-point model  
one can easily verify that the Atiyah-Singer argument leads to a result in complete agreement with equations \eqref{8.11} and \eqref{8.12}. 

To sum up, in the case of the triple-point semimetals model with lagrangian \eqref{2.13}, when the vector gauge invariance  is maintained,  the axial anomaly is given by 
\beq
 \partial_{\mu} J_A^{\mu} (x)=    \frac{\cos (3 \theta)}{| \cos (3 \theta) | } \,  \frac{1}{4 \pi^2} \, \epsilon^{\mu \nu \tau \lambda} \der_\mu  A_\nu (x) \der_\tau A_\lambda (x) \; .  
 \label{8.13}
\eeq
This concludes the derivation of equation \eqref{2.15}. {\color{black}  Let us recall that the axial anomaly \eqref{8.13} has been obtained for vanishing chemical potential; in our notations, this means that all the one-particle states with zero energy are assumed to be occupied.  A discussion on the stability of the result \eqref{8.13} against perturbations of the chemical potentials is contained in section~\ref{Stability}, as well as on its relation with the presence of semi-chiral states with asymptotically  vanishing energy \cite{bradlyn2016,ezawa17}. These semi-chiral states could also contribute  to the normalization of the vector current associated to the chiral magnetic effect \cite{burkov2012bis,burkov2012}, which is strictly related to the axial anomaly.   }

\section{Quantization of the anomaly coefficient}
\label{quantization}

In section~\ref{general}, it has been mentioned that the overall multiplicative factor $\cal N$, which appears in equation (\ref{3.10}), can only assume integer values. According to the    interpretations of Nielsen-Ninomiya and Atiyah-Singer of the axial anomaly \cite{nielsen1983,AS,Nakahara},    the quantisation of $\cal N$ naturally emerges. We would like to present here another argument ---confirming the quantisation of $\cal N$--- which is based on perturbation theory, and which is of  interest  because it makes use of  the relationship between the abelian and the nonabelian gauge anomalies \cite{BAZ,BA,ZWZ,BARZUM}.  
In the case of non-abelian anomalies, the quantisation of the anomaly normalization factor has been discussed for instance in \cite{INI,POL,THF}. 

Let us briefly recall where the emergence of the chiral gauge anomaly is found in perturbation theory.  Suppose that the lagrangian for a  fermion field $\psi_R (x)$ in the presence of a classical gauge potential $V_\mu(x)$ takes the form 
 \be
{\cal L}_R = \psi_R^\dagger (x) \, \Pi_R (V) \,  \psi_R (x) \; , 
\label{9.1}
 \ee
 in which $\Pi_R (V)$ represents a certain differential operator which is a function of the covariant derivatives $D_\mu = \der_\mu + i V_\mu $. As we have seen in the previous sections, the field $\psi_R(x)$ may contain several components, and  not necessarily it represents a spinor field.   Let us assume that ${\cal L}_R$ is invariant under local gauge transformations $\psi_R (x) \rightarrow  e^{i \theta_R (x)}  \psi_R (x) $, $V_\mu (x) \rightarrow  V_\mu (x) - \der_\mu \theta_R (x) $. The renormalized sum of the connected one-loop vacuum-to-vacuum diagrams of the $\psi_R(x)$ field ---in the presence of the classical background $V_\mu(x)$---  is denoted by $i \Gamma_R[V]$. Here it is assumed that  
 $\Gamma_R[V]$ admits a perturbative expansion in powers of the background gauge  field $V_\mu (x)$.    Despite the gauge invariance of ${\cal L}_R$, the infinitesimal gauge variation of $\Gamma_R [V]$ may be nonvanishing and, modulo the variation of local counterterms, it is given by  
 \be
\delta_{\theta_R} \Gamma_R[V] =  - {\cal N} {1\over 24 \pi^2} \int d^4 x  \, \epsilon^{\mu \nu \tau \lambda}   \der_\mu \theta_R (x)  V_\nu (x) \der_\tau V_\lambda (x)  \; .  
\label{9.2}
\ee
When ${\cal N}\not= 0$, the classical  gauge invariance $V_\mu (x) \rightarrow  V_\mu (x) - \der_\mu \theta_R (x)$ is broken by the presence of an anomaly.  We have already mentioned the universality of the local function $\epsilon^{\mu \nu \tau \lambda}   \der_\mu    V_\nu  \der_\tau V_\lambda  $ entering expression \eqref{9.2}. We would like to elaborate now on the quantisation of $\cal N$. 
 
 Let us consider the  chiral anomaly in the case of  a non-abelian gauge symmetry.  The field theory model defined by the lagrangian \eqref{9.1} will now be modified in order to introduce a non-abelian symmetry. 
  Suppose  that a certain field theory  model  contains $N$ (with $N > 2$) copies of the fermion field $\psi_R(x)$. Thus the variables of this new  model can be described by the fields $\psi_R^j (x)$, where the index $j$, that we call the flavour index, takes values   $j=1,2...,N$; this set of fields  will be  denoted by $\Psi_R(x)$. An internal symmetry group  acts on the $N$ components $\psi_R^j(x)$ of $\Psi_R(x)$ according to the fundamental representation  of $SU(N)_R$. Let now the gauge field $V_\mu (x)$  take values in the Lie algebra of $SU(N)_R$,  $V_\mu (x) = V_\mu^a (x) T^a $, where $\{ T^a \}$ (with $a=1,2,..., N^2-1$)  are the generators   of $SU(N)_R$.  Let the lagrangian of the model be  
  \be
{\cal L}_R = \Psi_R^\dagger (x) \, \Pi_R (V) \,  \Psi_R (x) \; , 
\label{9.3}
 \ee
 where, in the differential operator $\Pi_R (V)$,   the abelian covariant derivative $D_\mu = \der_\mu + i V_\mu (x)$  has been replaced by  
 the non-abelian covariant derivative $ \left ( D_\mu \right )_{jk} = \delta_{jk} \der_\mu + i V_\mu^a (x) T^a_{jk} $, and a sum over all the flavour indices is understood.  The lagrangian \eqref{9.3} is invariant under $SU(N)_R$  gauge  transformations. Under an infinitesimal $SU(N)_R$ gauge transformation  $V_\mu (x) \rightarrow V_\mu (x) - \der_\mu \theta_R (x)+ i [  \theta_R (x), V_\mu (x)] $, the gauge variation of the renormalized one-loop functional  $\Gamma^\prime_R[V] $  of the nonabelian model is given by 
 \be
\delta_{\theta_R} \Gamma^\prime_R[V] =  - {\cal N} {1\over 24 \pi^2} \int d^4 x  \, \epsilon^{\mu \nu \tau \lambda} \, {\rm Tr} \Bigl [  \der_\mu \theta_R (x)  \bigl ( V_\nu (x) \der_\tau V_\lambda (x) + (i/2) V_\mu (x) V_\nu (x) V_\lambda (x)\bigr ) \Bigr ]  \; .   
\label{9.4}
\ee
The fields polynomial which must be integrated in expression  \eqref{9.4} satisfies \cite{WZ} the Wess-Zumino consistency conditions. It is important to note that the multiplicative coefficient $\cal N$ that appears in equation \eqref{9.4} is exactly the same coefficient $\cal N$ entering equation \eqref{9.2}. This equality is well known in the context of the  computation \cite{BA,ZWZ} of the chiral anomalies by means of perturbation theory. 
Indeed,   in the perturbative computation of the anomaly (\ref{9.4}), the Feynman diagrams which contribute to the term of  expression (\ref{9.4}) which is quadratic in $V_\mu(x)$  precisely  coincide with the diagrams which enter the computation of the  abelian anomaly (\ref{9.2}).  In the perturbative computation of this term, the non commutativity of the $SU(N)$ generators is harmless  because the field combination $\epsilon^{\mu \nu \tau \lambda} \der_\mu  V_\nu  \der_\tau V_\lambda $ is symmetric under the exchange $\der_\mu  V_\nu   \leftrightarrow \der_\tau V_\lambda $.  
As far as these Feynman diagrams are concerned, the only difference between the abelian and the nonabelian case is that, in the nonabelian case, in the end of the computations one has to take a sum over the flavour indices, or, one needs to introduce a trace over the indices of the fundamental $SU(N)_R $ representation. Note that the presence of this trace is explicitly   indicated in expression \eqref{9.4}. Therefore the same Feynman diagrams  which produce the coefficient ${\cal N} $  in equation (\ref{9.2}) necessarily yield the same coefficient $\cal N$ in equation (\ref{9.4}). 
   
 At this point, in order to complete the argument,  we need to show that the coefficient $\cal N$ multiplying  the nonabelian anomaly \eqref{8.4} must take integer values. 
 Instead of displaying  a formal proof, let us produce a physical argument.  
   
The relationship of the abelian chiral anomaly \eqref{3.10} and the corresponding abelian axial anomaly \eqref{3.21} has been discussed in section~\ref{general}; let us consider the non-abelian generalisation of this relationship. In the nonabelian case, suppose that, in addition to the field $\Psi_R(x)$,  one also has the fermionic field $\Psi_L(x)$ made of  $N$ components $\psi^j_L(x)$, with $j=1,2,..,N$ and  the corresponding lagrangian term is 
  \be
{\cal L}_L = \Psi_L^\dagger (x) \, \Pi_L (W) \,  \Psi_L (x) \; . 
\label{9.5}
 \ee
The differential operator $\Pi_L(W)$ is a function of the covariant derivative $ \left ( D_\mu \right )_{jk} = \delta_{jk} \der_\mu + i W_\mu^a (x) T^a_{jk} $, where $W_\mu (x) = W_\mu^a (x) T^a $ is the connection of the  gauge group $SU(N)_L$, which acts on the components $\psi^j_L(x)$.  It is assumed that the lagrangian \eqref{9.5} is invariant under $SU(N)_L$ gauge transformations,  with $\Psi_L(x)$ transforming according to the fundamental $SU(N)_L$ representation. Let us assume that the chiral $SU(N)_L$ anomaly takes the form 
 \be
\delta_{\theta_L} \Gamma^\prime_L[W] =   {\cal N} {1\over 24 \pi^2} \int d^4 x  \, \epsilon^{\mu \nu \tau \lambda} \, {\rm Tr} \Bigl [  \der_\mu \theta_L (x)  \bigl ( W_\nu (x) \der_\tau W_\lambda (x) + (i/2) W_\mu (x) W_\nu (x) W_\lambda (x)\bigr ) \Bigr ]  \; ,    
\label{9.6}
\ee
where the multiplicative factor $\cal N$ is the same factor $\cal N$ appearing in expression \eqref{9.4}. This is precisely what one finds with fermion multiplets  (as quarks and leptons fields) of ordinary spinor fields $\Psi_R(x)$ and $\Psi_L(x)$ minimally coupled with gauge fields. 

In the  composed field theory model which contains  both the $N$-components field $\Psi_R(x)$ and the $N$-components field $\Psi_L(x)$ and total lagrangian 
\be
{\cal L} = \Psi_R^\dagger (x) \, \Pi_R (V) \,  \Psi_R (x) + \Psi_L^\dagger (x) \, \Pi_L (W) \,  \Psi_L (x) \; , 
\label{9.7}
\ee
where $V_\mu (x) $ and $W_\mu (x)$ are classical background fields,  one has an anomalous  flavour symmetry group $SU(N)_R \times SU(N)_L$. The $SU(N)_R \times SU(N)_L$ gauge variation of the one-loop functional $\Gamma^\prime [V,W]= \Gamma^\prime_R[V]  + \Gamma^\prime_L[W] $ is given by the sum of expressions \eqref{9.4} and \eqref{9.6}. Similarly to the abelian case, by adding a suitable finite local  Bardeen counterterm   $L_B[V,W]$ to the one-loop functional $\Gamma^\prime [V,W]$, one can define \cite{BA}  a new functional $\Gamma [V,W] = \Gamma^\prime [V,W] + L_B[V,W]$ which is invariant under transformations of the vector subgroup $SU(N)_V$ of  $SU(N)_R \times SU(N)_L$. Only axial transformations (with infinitesimal parameters given by the difference $\theta_R - \theta_L$) are anomalous. The resulting Bardeen flavour anomaly \cite{BA} of the axial component of the   group $SU(N)_R \times SU(N)_L$ is proportional to $\cal N$. 

 When this flavour  anomaly   is integrated \cite{WZ} by employing  for instance the so-called ``Goldstone bosons''  field \cite{CWZ,CCWZ} $U(x) \in SU(N)$, the corresponding Wess-Zumino term is proportional to ${\cal N}$.    The Wess-Zumino term is well defined \cite{WZ,NOV,W}  ---and the ambiguities which are originated  by  the obstruction given by the non triviality of  $\pi_5 (SU(N)) $  are harmless---    only when ${\cal N}  \in {\mathbb Z}$. On the other hand, the Wess-Zumino term  must be well defined because, in the case of the low energy  effective lagrangian \cite{CWZ,CCWZ} of the hadrons physics in which $N=3$, for instance, it describes part of the hadronic interactions of the light pseudo scalar mesons of the octet ---and part of the interactions between these mesons and the flavour gauge fields of the Standard Model---  which  can be observed in laboratory. A few consequences of the quantisation of $\cal N$ in  particles physics can also be found, for instance,  in references  \cite{INI,POL,THF,CDG,JR,BHO,SW81,EKO,RUB,WIL,EG84,CRE}.  Therefore the value of the multiplicative factor ${\cal N}$, entering expressions \eqref{9.6}, \eqref{9.4} and \eqref{9.2}  must be an integer.
   
The quantization of the anomaly multiplicative factor ${\cal N}$ has nontrivial consequences in the field  theory models in which the fermion lagrangian terms contain free parameters, as in the case of the parameters $\{ \alpha, \beta, \gamma \} $ in the multi-Weyl  models \eqref{2.5} and \eqref{2.11}. Indeed, since any smooth variation of a quantised coefficient must be vanishing, in these models the expression of the axial anomaly  must be  invariant under ``smooth''  variations of the parameters, {\it i.e.}  ``smooth''  modification   of the  operator acting on the fermion fields in  the lagrangian.    And  in facts, this is precisely the outcome of the explicit computations of the axial anomalies \eqref{2.9} and \eqref{2.12}, which depend on $\{ \alpha, \beta, \gamma \} $ through the variable 
$$
\Theta (\alpha , \beta , \gamma )  = {\frac{\alpha \beta \gamma }{| \alpha \beta \gamma |}} \; .
$$
The function $\Theta (\alpha , \beta , \gamma )$ is locally constant.  All the  modifications of the value of $\Theta (\alpha , \beta , \gamma )$ are found  when one of the parameters $\{ \alpha, \beta, \gamma \} $ changes its sign; that is, when the value of one of the parameters crosses the zero point. Note that  the zero value of one of these parameters represents a critical point for the lagrangian operators $\Pi_R(V)$ or $\Pi_L(W)$  entering the lagrangian. Indeed, when $\alpha = 0 $ or $\beta =0 $, for instance, there are no more normalizable solutions of equations \eqref{5.8} and \eqref{7.3}.  Consequently, a modification of the sign of one of the parameters $\{ \alpha, \beta, \gamma \} $ does not corresponds to a  ``smooth'' modification   of the operators $\Pi_R(V)$ or $\Pi_L(W)$.  

 Similarly, in the case of the triple-point semimetals model \eqref{2.13} the dependence of the axial anomaly \eqref{2.15}  on the $\theta$ parameter is given by the multiplicative factor   
$$
 \frac{\cos (3 \theta)}{| \cos (3 \theta) | } \; ,  
$$
 which is locally constant. The change of sign of this factor occurs for $3 \theta = \pm \pi /2 $, which correspond to critical points for the operators appearing in the lagrangian \eqref{2.13}. Indeed, as it has been shown in section~\ref{3-point}, when $3 \theta = \pm \pi /2 $ the zero eigenvectors of the reduced hamiltonian \eqref{8.4} are not normalizable. \\

A consequence of the quantization of the anomaly is that, if a term 
$\delta E ({\bf k}) \propto q \,  k_x \,  \big(\psi_{R}^\dag ({\bf k}) \psi_{R} ({\bf k}) \pm \psi_{L}^\dag ({\bf k}) \psi_{L} ({\bf k}) \big)$
 (terms proportional to $ \propto q \,  k_y$ or $ \propto q \,  k_z$ are possible as well) is added to the Weyl lagrangian \eqref{2.1}, no variation of the anomaly from the form \eqref{2.4} is obtained, until $q$ is strong enough to induce a transition to a type-II Weyl-semimetal \cite{soluyanov2015}. In the latter condition, the Fermi surface becomes extended and important deviations in the anomaly are expected. We predict the same situation for type-II generalizations of double- and triple-Weyl semimetals, driven by  terms as 
 $\delta E ({\bf k}) \propto q \, k_x^l \,  \big(\psi_{R}^\dag ({\bf k}) \psi_{R} ({\bf k}) \pm \psi_{L}^\dag ({\bf k}) \psi_{L} ({\bf k}) \big)$, with $l = 2,3$ respectively.

{\color{black}
\section{Chemical potentials and stability}
\label{Stability}

The computations of the axial anomaly that have been presented in the previous sections refer to the case in which all the single-particle states with negative energy  are occupied. This corresponds to the situation   in which the chemical potential coincides with the energy of the  band-touching nodes of the semimetals. Let us now consider the stability of our results under modifications of the chemical potentials. 

\subsection{Multi-Weyl semimetals}

In the perturbative approach, a discussion on the stability of the chiral gauge anomalies for a single Weyl model can also be found, for instance, in the paper \cite{BA} by Bardeen, in which the most general bilinear couplings  of the fermions with external sources have been considered. Actually, the stability of the axial anomaly has a general validity that we shall now examine. 

As it is shown in equations \eqref{2.5}, \eqref{2.11} and \eqref{2.13}, the lagrangian density $\cal L$ of the various models that we have considered in the present article has the common structure
\be
{\cal L} = \Psi^\dagger (x)\, \left \{ i D_0 - {\cal H}(\bm D) \, \right \} \Psi (x) = \Psi^\dagger (x)\, \left \{ i \left (\der_0 + i A_0 \right )- {\cal H}(\bm D) \, \right \} \Psi (x) \; , 
\label{11.1}
\ee 
where $\Psi = \left ( \psi_R , \psi_L\right )$ and ${\cal H}(\bm D)$  represents a differential operator constructed with the spatial components of the covariant derivative  $ \bm D_j = \der_j + i A_j (x)$, with $j=1,2,3$.    The modification of the chemical potentials can be described by the introduction of  the additional lagrangian term 
\be
\Delta {\cal L} = \mu_R \,  \psi_R^\dagger (x) \psi_R(x) +  \mu_L \, \psi_L^\dagger (x) \psi_L(x) \; , 
\ee
where $\mu_R$ and $\mu_L$ are constant parameters. For sufficiently small values of $\mu_R$ and $\mu_L$, does the addition of  $\Delta {\cal L} \not=0$ modify the expression of the axial anomaly ? In other words, is the axial anomaly, computed for the model which is described by the lagrangian ${\cal L} + \Delta {\cal L}$, equal to the axial anomaly which is found for the model with lagrangian $\cal L$~?   

When $\mu_R = \mu_L =  \mu $, the modified lagrangian density ${\cal L} + \Delta {\cal L}$ takes the form 
\be
{\cal L} +\Delta {\cal L} =  \Psi^\dagger \, \left \{ i \left [ \der_0 + i (A_0 - \mu ) \right ]- {\cal H}(\bm D) \, \right \} \Psi  \; , 
\ee
which is equal to expression \eqref{11.1} with the  only replacement $A_0 (x) \rightarrow A_0(x) - \mu$. The axial anomaly expression  $\epsilon^{\mu \nu \tau \lambda} \der_\mu A_\nu (x) \der_\tau A_\lambda (x)$ is not modified by 
the replacement $A_0 (x) \rightarrow A_0(x) - \mu$. Therefore the introduction of the chemical potential $\mu \not= 0$ does not alter the expression of the axial anomaly. The stability of the axial anomaly if  $\mu_R = \mu_L$  is suggested directly also by the  Nielsen-Ninomiya method.    
Indeed, the addition of a term  $\mu \, \psi^\dagger (x) \psi (x)$  modifies the zero-point of the energy spectrum but, in the presence of electric and magnetic fields, it does not modify the crossing rate of the energy values through the zero level.  Therefore, the introduction of $\mu$ would simply amount to a constant energy shift $\omega \to \omega +\mu$ which does not affect the values of the rates $\der_0 N_R$ and $\der_0 N_L$ of equations \eqref{5.24} and \eqref{5.26}.

When $  \mu_L = - \mu_R = \mu_5$, the  lagrangian density ${\cal L} + \Delta {\cal L}$ takes the form 
\be
{\cal L} +\Delta {\cal L} =  \Psi^\dagger \, \left \{ i \left [ \der_0 + i A_0 + i \mu_5 \gamma^5  \right ] - {\cal H}(\bm D) \, \right \} \Psi  \; , 
\ee
where $\gamma^5 = \hbox{diag} (1,-1)$. 
This expression describes the lagrangian of a  model in which  the fermion variable $\Psi (x)$ is coupled with the gauge connection $A_\mu $ of the group $U(1)_V$ in the usual way, and $\Psi (x)$  is also coupled with the gauge connection $B_\mu (x)$  of the group $U(1)_A$  in which $B_0(x) = \mu_5  $ and  $\bm B (x) =0 $.   As it has been demonstrated in Section \ref{general}, in this case, when the $U(1)_V$ gauge invariance is preserved, the axial anomaly is given by expression \eqref{3.20} 
\be
\delta_{\theta_A} \Gamma [A,B] =  {\cal N} {1\over 12 \pi^2 } \int d^4 x  \, \epsilon^{\mu \nu \tau \lambda}    \theta_A (x) \left [  \der_\mu B_\nu (x) \der_\tau B_\lambda (x) + 3 \der_\mu A_\nu (x) \der_\tau A_\lambda (x)  \right ] \; . 
\label{11.5}
\ee 
This equation shows that the contribution of the $B_\mu$ field to the axial anomaly is vanishing when $B_0(x) =0 $ and $\bm B (x) =0 $, and it is vanishing also when  $B_0 (x)= \mu_5  $ and  $\bm B (x) =0 $.  
Therefore,  the introduction of the chemical potential $\mu_5 \not= 0$ also does not modify the expression of the axial anomaly.  Note that the presence of $\mu_5 \not= 0$ gives rise to the  chiral magnetic effect \cite{burkov2012,burkov2012bis}, which is  connected to the axial anomaly and  is described by the vector current
\begin{equation}
j_\nu = - {\cal N} \, \mu_5 \,  \frac{e^2}{2\pi^2}   \epsilon^{0\nu \alpha \beta}\partial_\alpha A_\beta \, .
\end{equation}

\subsection{Triple-point semimetals}

The computation of the axial anomaly and the stability of the corresponding result in the case of the triple-point semimetal deserves a particular discussion. Let us recall that the lagrangians of the type shown in equations  \eqref{2.5}, \eqref{2.11} and \eqref{2.13}  represent phenomenological approximations of more  complicated  theories which describe the dynamics of the relevant fermionic degrees of freedom in the various materials.  Usually, the  validity of these simplified expressions is limited to  a neighbourhood of the locations of the touching nodes in the  Brillouin zone; in our notations, these neighbourhoods correspond to the low momenta regions. 
The study of the effective models, which are defined by the phenomenological lagrangians,  may by useful because, in certain cases, one can easily deduce interesting  features which are common to both   these low-energy models and to the true  physical systems.  The computation of the axial anomaly is precisely one example of this  strategy, in which it is assumed that the axial anomaly of the effective theories coincides with the axial anomaly of the corresponding real systems.  However, it turns out that the low-energy model associated with the triple-point semimetal, which is defined by the lagrangian \eqref{2.13}, presents certain  unrealistic peculiarities that must be taken into account in order to determine the axial anomaly.

In the case of vanishing gauge  potential, $A_\mu(x)= 0$, the energy spectrum of the one-particle states, which is defined by the lagrangian \eqref{2.13} when $\theta = 0$, for instance,  contains a  flat band with vanishing energy, such that $\omega ({\bm k}) =0 $, for all values of $\bm k$.  As a consequence, standard perturbation theory  --- in which one makes an expansion  of the Feynman diagrams in powers of the classical fields $A_\mu (x)$ --- cannot be defined for  the theory \eqref{2.13}, because the Feynman propagator does not exist. The existence of a flat band of this type, with arbitrarily large momentum and zero energy, is usually unstable under realistic perturbations of the Hamiltonian and can be considered as an unphysical artifact of the model.

 Although the Feynman diagrams approach  cannot be utilised, in order to determine the axial anomaly, one can still use the Nielsen-Ninomiya method (or, equivalently, the Atiyah-Singer approach), as it has been illustrated in Section~\ref{3-point}.  
But also in this case one finds certain unphysical features that must be taken into account. Indeed, as it has been shown in 
\cite{bradlyn2016,ezawa17}, in the presence of a magnetic field directed, for instance, along the $x^3$ direction, in addition to the chiral states with wave functions \eqref{8.6},  other two  Landau levels of normalizable states (here quoted ``semi-chiral") emerge, which have  zero energy only asymptotically ($k_3 \to \pm \infty$); for fixed chirality, in appropriate simplified notations, the corresponding dispersion relations have the hyperbolic  form $\widetilde \omega^{(\pm)} (k_3) = k_3  \pm \sqrt {k_3^2 + B^2}$. The branch $\widetilde \omega^{(+)} (k_3)$ describes particle  states with decreasing and vanishing energy as $k_3 \rightarrow - \infty$, whereas $\widetilde \omega^{(-)} (k_3)$ corresponds to antiparticle (or hole) states with decreasing and vanishing energy as $k_3 \rightarrow \infty$.  
 
 The asymptotic behaviour of $\widetilde \omega^{(\pm)} (k_3) $ in the large $|k_3|$ limit  appears to be rather unreliable.  
 Indeed, for realistic lattice (tight-binding) models, it is likely that the semi-chiral states display a modified dispersion at momenta sufficiently far from the nodes, and a finite separation $\Delta$ in energy from the zero level, due at least to the finite extension of the Brillouin zone.  Moreover, these semi-chiral states  could be subject to a strong mixing effect due to the magnetic field. Their peculiar dispersion which approaches zero energy at large momenta $k_3$  can indeed favor the coupling of the hyperbolic branches belonging to triple-point crossings with opposite chiralities and we reckon the consequent mixing effect to be stronger than the one for Weyl semimetals \cite{kim2017}. In this scenario, the semi-chiral modes would develop a finite energy gap, separating them from the flat band. 

Let  us now concentrate on the two chiral Landau levels that appear for positive energies and are relevant for the  computation of the axial anomaly.
The first  has a dispersion relation $\omega_1 (k_3) $ linear in $k_3$ (in appropriate notations one can put $\omega_1(k_3) = 2 k_3$), the wave functions of the corresponding states are shown in equation \eqref{8.6}. The second   has instead a dispersion relation of the hyperbolic form $\omega_2 (k_3) = \widetilde \omega^{(+)} (k_3) = k_3  + \sqrt {k_3^2 + B^2}$.   The  behaviours of $\omega_1(k_3)$ and $\omega_2 (k_3) $ are sketched in Figure \ref{fig:landau}. 
\begin{figure}[ht]
\begin{tikzpicture} [scale=0.8] [>=latex]
\draw [thick, -> ]  (-5,0) -- (5,0);
\draw [thick, -> ]  (0,-2) -- (0,5); 
\draw [very thick ]  (-2,-2) -- (4.5,4.55); 
\draw  [very thick ]  (-4.9,0.35) .. controls (-1,0.57) and (2,2.6) .. (4,4.4); 
\draw [very thick,  dashed  ]  (-4.9,.5) -- (4.8,.5);
\node at (-.3, 4.6){$\omega $}; 
\node at (4.6, -0.36){$ k_3 $}; 
\node at (0.3, -0.3){$0 $};
\node at (3.2,2.56) {$  \omega_1  $};
\node at (-1.4,1.5) {$  \omega_2  $};
\node at (3,0.77) {$  \mu  $};
\end{tikzpicture}
\caption{Energy levels $\omega_1(k_3)$ and $\omega_2(k_3)$ for single-particle states,  obtained from the effective theory of equation  \eqref{2.13}.}
 \label{fig:landau}
\end{figure}
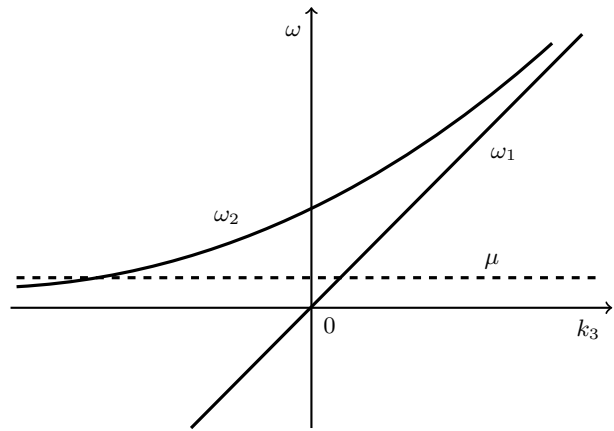

For vanishing chemical potential, only the branch $\omega_1 (k_3)$ intersects the zero energy level and, according to the Nielsen-Ninomiya procedure,  it gives a unitary contribution to the integer coefficient $\cal N$ appearing in  the axial anomaly.  However, as it is shown in Figure~2, both energy branches $\omega_1 (k_3)$ and $\omega_2 (k_3)$
intersect the energy level determined by a nonvanishing value $\mu > 0 $ of the chemical potential. So, one could claim that, in this case,  the multiplicative integer coefficient of the axial anomaly must be doubled, $|{\cal N}|=2$ (with a negative value of the chemical potential, one needs to consider the branch $\widetilde\omega^{(-)} (k_3)$ instead of $\widetilde\omega^{(+)} (k_3)$, getting the same conclusion).  However, the value of $k_3$ corresponding to the intersection point of $\omega_2 (k_3)$ with the $\mu $ energy level  tends to $- \infty $ as $\mu $ approaches to zero. Thus, for sufficiently small $\mu$, the  intersection point is outside the validity range of the   low-energy effective theory \eqref{2.13}, and the lattice corrections mentioned above  should no more be  neglected.  This is why, in the computation of the axial anomaly for the triple-point semimetals presented in Section~\ref{3-point},  we have chosen  to take into account of the dispersion relation $\omega_1 (k_3)$  exclusively, thus modeling the behavior for $\mu$ approaching zero. 

Summing up, we expect that our result $|{\cal N}| =1$ for the axial anomaly coefficient of the triple-point semimetals
is stable under sufficiently small modifications of the chemical potentials from zero energy. 
 If the mixing of the semi-chiral modes due to the magnetic field is not strong enough 
to develop a gap, by varying the chemical potential of the system, we possibly expect to find
a transition from a phase displaying $|{\cal N}| =1$ for small $\mu$, to a phase 
with $|{\cal N} |=2$ for larger values of it. 
}

\section{Conclusions}  
\label{conclus}
In this article we have derived the expression of the abelian axial anomaly for the double-Weyl, triple-Weyl and triple-point semimetal models. 
Three different computation methods have been considered: the perturbative quantum field theory procedure which is based on the evaluation of the one-loop Feynman diagrams,  the Nielsen-Ninomiya  method, and  the Atiyah-Singer index argument.  The consistency of these methods,  which have been  shown to be closely related, has been illustrated in detail in the case of the double-Weyl model. 
For the triple-Weyl and the triple-point  models the perturbative approach is rather burdensome or affected by singularities; therefore only the Nielsen-Ninomiya and Atiyah-Singer methods have been discussed.
It has been shown that the   dependence of the anomaly on the vector gauge field $A_\mu (x)$  is not contingent on the Lorentz symmetry,  but is determined by the gauge symmetry structure. In facts,   the axial anomaly takes the general form 
$$
 \partial_{\mu} J_A^{\mu} =   {\cal N} \frac{1}{16 \pi^2} F \wedge F \; , 
$$
where $F =  ( \der_\mu A_\nu - \der_\nu A_\mu ) \,  dx^\mu \wedge dx^\nu $ denotes the curvature 2-form, and the value of the multiplicative factor $\cal N $ is determined by the lagrangian of each model. 
   General arguments,  still based on gauge invariance, suggest  that the factor $\cal N$ must be quantized and must match the topological charge of the corresponding band touching points. Indeed, this is precisely the outcome of the explicit anomaly computations. 
 We have found that $|{\cal N}| = 2 $ for the double-Weyl model, $|{\cal N}| = 3 $ for the triple-Weyl model and $|{\cal N}| = 1 $ for the triple-point model. The last result has been obtained by neglecting the hyperbolic Landau levels with asymptotical vanishing  energy in the limit of large momenta. Indeed, we have presented arguments supporting this choice.   For this reason, our result is not in contradiction with the usual counting of the chiral states in the triple-point semimetals presented in \cite{bradlyn2016,ezawa17}).   We have further discussed the stability of the anomaly under smooth modifications of the lagrangian parameters,  showing that the value of $\cal N$ is invariant under these deformations.
 The modification of the sign of $\cal N$ in the considered models has been examined.  We have verified that,  in the parameter space, the points  in which the value of $\cal N$ undergoes a  change of sign indeed correspond to critical points.
{\color{black}  Finally we have shown that, in agreement with the case of a single-Weyl model, a modification of the chemical potentials does not change the expression of the axial anomaly.}

\section*{Acknowledgements}
It is a great pleasure to thank Ion Cosma Fulga, Massimo Mannarelli,  Michele Mintchev, Giampiero Paffuti, Simone Paganelli, and Andrea Trombettoni for useful discussions.

\end{document}